\documentclass[11pt,a4paper]{article}

\usepackage[
top    = 2.50cm,
bottom = 2.50cm,
left   = 2.50cm,
right  = 2.50cm]{geometry}
\usepackage{jheppub}
\usepackage[sort&compress]{natbib}

\usepackage{amssymb}
\usepackage{graphicx}
\usepackage{slashed}
\usepackage{amsmath}
\usepackage{hyperref}
\usepackage{verbatim}

\usepackage{caption}
\usepackage{subcaption}
\usepackage{array,tikz-cd}
\usetikzlibrary{calc}

\newcommand{\Tr}{\mbox{Tr}}

\title{Chord diagrams, exact correlators in spin glasses and black hole bulk reconstruction}

\author[a]{Micha Berkooz}
\author[b]{\!, Prithvi Narayan}
\author[c]{\!, Joan Sim\'on}
\affiliation[\,a]{
\it{Department of Particle Physics and Astrophysics,\\
Weizmann Institute of Science, Rehovot 7610001, Israel}}
\affiliation[\,b]{
\it{Department of Physics, Indian Institute of Technology, Palakkad 678557, India}}
\affiliation[\,c]{
\it{School of Mathematics and Maxwell Institute for Mathematical Sciences,\\
        University of Edinburgh,
        Edinburgh EH9 3FD, United Kingdom}}
\emailAdd{Micha.Berkooz@weizmann.ac.il}
\emailAdd{prithvi.narayan@gmail.com}
\emailAdd{j.simon@ed.ac.uk}

\abstract{The exact 2-point function of certain physically motivated operators in SYK-like spin glass models is computed, bypassing the Schwinger-Dyson equations. The models possess an IR low energy conformal window, but our results are exact at all time scales. The main tool developed is a new approach to the combinatorics of chord diagrams, allowing to rewrite the spin glass system using an auxiliary Hilbert space, and Hamiltonian, built on the space of open chord diagrams. We argue the latter can be interpreted as the bulk description and that it reduces to the Schwarzian action in the low energy limit.}

\begin{document}

\maketitle

\section{Introduction}

Over the last few years the Sachdev-Ye-Kitaev (SYK) model \cite{Sachdev:1992fk,KitaevTalks,Maldacena:2016hyu,Polchinski:2016xgd,Sachdev:2010um} has attracted considerable attention as a model of quantum qravity which exhibits the correct chaotic behaviour. The SYK model at large $N$ turns out to be solvable, in the IR and other limits, and exhibits several interesting features like conformal regime \cite{Sachdev:1992fk} and maximal chaos exponent \cite{KitaevTalks,Polchinski:2016xgd,Maldacena:2016hyu,Maldacena:2015waa}. Its pattern of symmetry breaking is encoded by the Schwarzian low energy effective action \cite{Maldacena:2016hyu, Jensen:2016pah,Maldacena:2016upp}, which is the effective dynamics of gravity on $AdS_2$, coupled to a scalar field \cite{Almheiri:2014cka}.

There are many interesting generalizations of this model, starting from the original work of \cite{Sachdev:1992fk}, like those in higher dimensions \cite{1999PhRvB..59.5341P,Berkooz:2016cvq,Gu:2016oyy,Murugan:2017eto,Berkooz:2017efq}. Models without disorder with SYK like physics have also been proposed in \cite{Witten:2016iux, Klebanov:2016xxf,Gurau:2010ba}. The long time scales of the SYK model have been discussed and connections to random matrix theory pointed out in \cite{Cotler:2016fpe,Gharibyan:2018jrp}. Higher point functions in the model have been computed in \cite{Gross:2017hcz,Gross:2017aos}. For work beyond large $N$, see \cite{Garcia-Garcia:2016mno,a:2018kvh,Garcia-Garcia:2017pzl}. The near $AdS_2$ spacetime   
interpretation was elaborated in \cite{Sachdev:2015efa,Mandal:2017thl,Jensen:2016pah,Cvetic:2016eiv,Engelsoy:2016xyb,kitaev2017soft,Das:2017hrt,Das:2017wae,Das:2017pif} (see \cite{Jevicki:2016bwu,Jevicki:2016ito} for related work on bilocal fields). 

In this work, we will be interested in a particular class of spin glass models introduced in \cite{Erdos}, which are close relatives of the SYK model, and derive a formula for the exact 2-pt function of certain operators. The model is the following. Consider $n$ sites with a spin $\frac{1}{2}$ degree of freedom on each.  Denote the Pauli matrices acting on site $i=1,2,\dots ,n$ by $\sigma^{(a)}_i$, with $a=1,2,3$. Given an integer p, we define a random Hamiltonian $H^{(p)}$ as follows. Let $e=(i_1,....i_p)$ be a vector of length $p$ of distinct integers defining a subset of the $n$ sites, and let $a=(a_1,...a_p)$ be a second vector of length $p$, with entries being either 1,2 or 3. Denoting the pair $(a,e)$ by J, we define
\begin{equation}\label{RandHam}
	\sigma_J=\sigma_{(a,e)}=\sigma_{i_1}^{(a_1)}\sigma_{i_2}^{(a_2)}....\sigma_{i_p}^{(a_p)}
\end{equation}
and the spin glass Hamiltonian is
\begin{equation}\label{Hamilt}
	H^{(p)}=3^{-p/2}{n\choose p}^{-1/2}\sum_J \alpha_J \sigma_J
\end{equation}
where the sum runs over all possible $J$'s, and $\alpha_J$ are independent Gaussian variables with zero mean and unit standard deviation (we will drop the superscript p from now on). The relevant parameter controlling the asymptotic density of states is \cite{Erdos}
\begin{equation}\label{ParamDef}
	q=e^{-\lambda} \quad \text{with} \quad \lambda= \frac{4}{3}\ \frac{p^2}{n} \ ,
\end{equation}
and the exact asymptotic density of states of the model \eqref{Hamilt} was computed in \cite{Erdos} in the limit 
\begin{equation}\label{eq:lambda scaling limit}
	\lambda \  \text{fixed}, \hspace{10mm }\ n\to\infty
\end{equation}
We will refer to this as the $\lambda$-scaling limit. We will be interested in the limit of $\lambda\to 0$, where the distribution of eigenvalues approaches a Gaussian distribution (point-wise) and hence we will refer to these models as  ``\textit{Almost Gaussian}" spin glass models. 

The spin glass model \eqref{Hamilt} is quite similar to the SYK model. Apart from replacing Majorana fermions with Pauli matrices, the more critical difference is that in the $\lambda$-scaling limit \eqref{eq:lambda scaling limit} $p$ is scaled with $\sqrt{n}$, leaving $\lambda$ as a parameter, whereas in SYK, $p$ is held fixed as $n\to\infty$, while scaling the energies properly to obtain a solution of the model. However, the $\lambda$-scaling model with Majorana fermions was discussed in \cite{Cotler:2016fpe} where it was shown to have a low energy limit $\lambda\to 0, E\to 0$, (dubbed ``double scaled SYK" model) where the density of states is that of the Schwarzian theory. Hence, models \eqref{Hamilt} can just as well be used to study the physics of $AdS_2$. In this work, we discuss the full \textit{Almost Gaussian} model and use the "double scaled limit" to check our results.

The main results in the paper are 
\begin{itemize}
\item We motivate why random operator observables are relevant for black hole physics, i.e., not just random Hamiltonians. This is done in section \ref{sec:Motivation and Summary}, where we also survey existing results and state the new result on the 2-pt function. 
\item A new method of computing the distribution of eigenvalues of the Hamiltonian \eqref{Hamilt} in the $\lambda$-scaling limit. The new method relies on the reduction in \cite{Erdos} of the spin glass Hamiltonian to chord diagrams but then takes a different route in evaluating the latter. This is done in section \ref{sec:new derivation}. Section \ref{sec:qeq1 limit} is an analysis of the $\lambda\to 0$ limit which parallels Appendix B of \cite{Cotler:2016fpe} in our notation.
\item The derivation in section \ref{sec:new derivation} relies on an auxiliary Hilbert space and a Hamiltonian acting on it, which we denote by $T$. This new Hamiltonian is equivalent\footnote{With one important exception : the trace is replaced by some choice of initial and final states.} to the full Hamiltonian of the spin glass in that whenever the unitary operator $e^{iHt}$ appears, acting on the original Hilbert space, it can be replaced by $e^{iTt}$ acting on the auxiliary Hilbert space. In section \ref{sec:Bulk} we suggest that this is the analogue of the bulk Hamiltonian and show in what limit it reduces to the Schwarzian effective action in its Liouville form.
\item In section \ref{sec:2 point function} we compute the exact time dependent 2-pt function of an additional random operator of length $p' \sim \sqrt n$. This can be reduced to another chord partition function in which one chord is marked. We use the technique developed in section \ref{sec:new derivation} to evaluate it. 
\end{itemize}


\section{Motivation, Setup and Summary of results}\label{sec:Motivation and Summary}

We will analyse the spin glass Hamiltonian model \eqref{RandHam}-\eqref{Hamilt}. However, we will probe it using a random operator. The latter will be of a similar statistical type as the Hamiltonian, i.e. it will be defined by the same equation \eqref{RandHam}-\eqref{Hamilt} but with 
\begin{itemize}
\item a different length parameter $p'\not=p$, and 
\item a new set of independently drawn coefficients.  
\end{itemize}
In subsection \ref{subsec:motivation} we motivate this specific choice of operator. The rest of the section is an ``executive summary" of the setup of the model and known results in \ref{subsec:known results}, and a summary of the new results in \ref{subsec:new results}.

\subsection{Motivation - random observables and factorization}
\label{subsec:motivation}

\subsubsection{Why random operator probes ?}

Since black holes share some properties with chaotic systems \cite{Sekino:2008he,Shenker:2013pqa,Maldacena:2013xja,Roberts:2014isa}, they can be thought of as described by a suitable random Hamiltonian. In particular, for AdS black holes, we might want to think about some core of states in the spectrum governed by a random Hamiltonian, describing the near horizon black hole physics, dressed by a ``structured" non-random Hamiltonian describing excitations well separated from the horizon. In this picture, one needs to specify the statistical class of the random Hamiltonian. This is precisely what the SYK model achieves, as the relevant class for nearly-AdS$_2$ spacetimes. 

The next step is to probe the black hole (BH) using the available bulk probes, such as single trace operators or their analogues. The Hamiltonian in quantum mechanics, or the local energy-momentum tensor in higher dimensions, is one such operator. Probing with the full Hamiltonian does not provide any more information beyond the partition function, but the local energy momentum operator does. In practice, it is another massless field for which we can put sources on the boundary. Just as the full Hamiltonian is indistinguishable from a random operator when acting on the BH states, we can expect that the local energy momentum operator will also be effectively described by some random (local) operator acting on the Hilbert space of the states of the black hole. 

But the local energy-momentum tensor is just one of a tower of single trace operators with which we can probe the system. In ${\cal N}=4$ SYM we can use its primary $\text{tr}(X^2)$ to probe the black hole, or we can just as well use any other of the $\text{tr}(X^n)$ operators. If the former is a random operator on the states of the black hole, why should we not expect that all single trace operators be of a similar nature ? {\it We would like to suggest that the relevant probes appearing in General Relativity are random operators on the BH states}\footnote{A similar suggestion was made in \cite{Balasubramanian:2014gla} for a different ensemble, and a related discussion for long time scales appears in \cite{Barbon:2014rma}.}. The main issue would then be from what ensemble these operators are drawn. If we have some idea about the statistical ensemble of the Hamiltonian, we can try and guess what is the ensemble for the other single trace operators.

Another way to phrase the argument is that the SYK model is dual to AdS$_2$ in an appropriate large N and energy regimes. But there are other models which realize the same universality class (for example, the one discussed in this paper is based on different spin matrices). So there may be many ways of defining the statistics of the random Hamiltonian which give rise to the same physics - some may be similar to SYK and others may be different. Focusing on the computation of specific operators used to define a specific realization, such as $\chi_i$ in the SYK model, certainly yields the maximal amount of information about the model but it may not be universal enough throughout the different models. Rather, motivated by the fact that the local energy-momentum tensor is a one "single trace operator" out of many, we would like to suggest that useful probes are random operators appropriately made out of the basic constituents of the theory, just like the Hamiltonian is. The statistical class of these random operators may be more universal throughout the different ways of building models (as we will see in our case).

Yet another argument is the following. In the SYK model, the Hamiltonian is a sum of finite rank polynomials of the $\chi$ fields with random couplings. Viewing the $\chi$'s as the analogues of the single trace operators in higher dimensions (which is anyhow problematic since they live in $\mathrm{SO(N)}$ representations) implies that the Hamiltonian in the black hole regime can be written as a sum of polynomials of single trace operators. This seems to be a very strong assumption for the higher dimensional AdS/CFT dualities. A weaker assumption is that both the local energy momentum tensor and all other single trace operators can be written using some other operators which act on the BH states, which are just used to define the statistical class of the operator and probes. These operators need not be asymptotic observables outside the black hole but rather they just need to be a rich enough set to allow for the correct definition of the statistical class of the observables.

This is somewhat against the usual application of the AdS/CFT correspondence where, in this context, the SYK model is taken to be the microscopic theory which defines all of spacetime. In this approach, one is committed to all the operators defined in the model. However, in practice if one is interested in the AdS$_2$ part, one glues it to an external region in order to break conformal invariance (and the gluing might eventually vary if, for example, one thinks about an AdS$_2$ near horizon of an object in higher dimensions). It is not clear to what extent the full SYK provides an extension which has an adequate gravity dual outside the AdS$_2$ region, and even if it does, it is not clear whether it is universal. The right probes on AdS$_2$ are determined just as much by this outside-of-AdS$_2$ region since the probe must be defined on the boundary. This means the choice of right probes in the AdS$_2$ region, within a given model, might be ambiguous in general.

\subsubsection{What random operator probes ?}

Having argued that random operators are suitable probes, with ensembles related to the one from which the Hamiltonian is taken, in this subsection we would like to discuss another constraint on the ensemble from which probes are drawn, originating from requiring factorization of correlation functions. We will see that it again points us in the direction of almost Gaussian random operators, similar to $H^{(p)}$.

Within the AdS/CFT correspondence correlation functions of single trace operators factorize at leading order when evaluated in the ground state or in any other state well described by a semiclassical background. This is usually taken to include black holes, although this assertion is on less solid footing there, as the detailed quantum state of the black hole may matter (and surely does over long enough time scales). So the extent to which correlation functions do not factorize will teach us about the role of the quantum state of the black hole, and may also teach us about deviations from the standard Einstein-Hilbert low energy effective action.

In the field theory side of the AdS/CFT correspondence, factorization is a consequence of the large N limit when evaluated around the ground state. Around the black hole background it implies a non-trivial constraint on the statistics of probe operators \cite{Balasubramanian:2014gla}. Consider a microcanonical ensemble with a small enough energy spread, and consider the 4-pt function\footnote{We will insert the operators at distinct but close enough times. The argument does not hinge on these details.} 
\begin{equation}\label{FourPtM}
	\text{Tr} \bigl( M M^\dagger M^\dagger M \bigr),
\end{equation}
where the trace is over the states in this energy band. Since black holes are strongly mixing systems, one might have expected that M - when acting in this energy band - will be described by one of the ordinary random matrix ensembles. An example of this is the often assumed strong form of ETH 
\begin{equation}\label{SETH}
	M_i^j {M^\dagger}_k^l \propto \delta_i^l \delta_k^j.
\end{equation}
where the most straightforward interpretation of this formula is as a statement about statistics of the matrix elements\footnote{We will assume that the operator has no 1-pt function, and in any case we can shift it away.}. This relation actually comes about by a minimal set of assumptions - that A) only pairwise contraction of the operators matter - after all, we would like to obtain factorization, and B) that all the states in the microcanonical energy band are equivalent and hence the statistics should have full unitary invariance in this energy band - this is also an assumption often made in statistical physics.

However, under these circumstances correlation functions do not factorize properly. The ansatz in \eqref{SETH} is the same as drawing the operator M from a distribution with measure
\begin{equation}
	e^{- {\cal N} Tr(MM^\dagger)},
\end{equation}
where ${\cal N}$ is the number of states in the energy band.
So we only need to compute a Gaussian integral.  With this measure, in the large ${\cal N}$ limit, the 4-pt function \eqref{FourPtM} receives only one (planar) contribution. However, factorization implies that there are two contractions. It seems difficult to remedy this within the ordinary ensembles (for example, by changing the measure to $e^{-{\cal N} V(M,M^\dagger)}$ for a more general $V$).

Since there are restrictions to implementing factorization in the simplest ensembles, it is interesting to find additional examples in which correlators factorize. More precisely we would like them to almost factorize - the deviation from exact factorization is then interpreted as bulk interactions. At the level of a single operator, the most naive indicator  of factorization - neglecting for the moment the issue of time dependence - is that 
\begin{equation}
	E\bigl( \langle M^{2k} \rangle \bigr) \sim  A^{2k}(2k-1)!!
\end{equation}
for an hermitian operator M, where $E()$ is the statistical average over the ensemble from which the operator is drawn (and $A$ is a factor set by the normalization of the operator). 

The ensembles in \cite{Erdos} are precisely of this type. For any operator of the form \eqref{RandHam}-\eqref{Hamilt}, the distribution of eigenvalues approaches the Gaussian one in the limit $\lambda\rightarrow 0$, so all operators with ${ p \over \sqrt{n} } \ll 1$ will be approximately Gaussian. If the Hamiltonian has a specific (small) $\lambda$, then operators for all other values of (small) $\lambda$ are in qualitatively a similar statistical class and approximately factorize. We will use them as our probes.

\subsection{Set up of the model and summary of known results}
\label{subsec:known results}

The model discussed in \cite{Erdos} is defined in equations \eqref{RandHam} to \eqref{eq:lambda scaling limit}. One of the main results in that paper is that the asymptotic distribution of eigenvalues, in the limit
\begin{equation}\label{scllimit}
	n\to\infty \hspace{20mm} \lambda  \text{ fixed},
\end{equation}
is given by (recall $q=e^{-\lambda}$)
\begin{equation}
	v(E|q)=\frac{\sqrt{1-q}}{\pi \sqrt{1-\frac{\sqrt{1-q}}{4}\,E^2}} \prod_{k=0}^\infty  \left[ {1-q^{2k+2}\over 1-q^{2k+1}} \left(  1-  {(1-q)q^k\over (1+q^k)^2 }\,E^2 \right)  \right]
\label{ESDist}
\end{equation}
in the range $E \in \left[-\frac{2}{\sqrt{1-e^{-\lambda}}},\,\frac{2}{\sqrt{1-e^{-\lambda}}}\right]$, and vanishes outside this region.

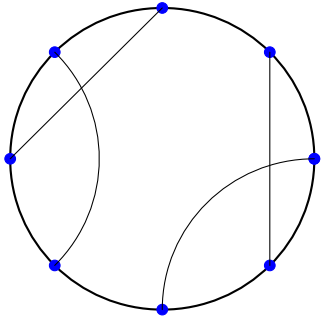
\begin{figure}[t!]
	\def\r{2}
	\begin{center}    
		\begin{tikzpicture}[scale=1,cap=round,>=latex]
		\draw[thick] (0,0) circle(\r);
		\foreach \x in {0,45,...,360} 
		{
			\filldraw[blue] (\x:\r) circle(2pt);
		}
		\draw[thin,color=black] (0,-\r) arc (180:90:\r) ;
		\draw[thin,color=black] (\r/1.414,\r/1.414) -- (\r/1.414,-\r/1.414);
		\draw[thin,color=black] (-\r/1.414,\r/1.414) arc (45:-45:\r) ;        
		\draw[thin,color=black] (-\r,0) -- (0,\r);
		\end{tikzpicture}
		\caption{A sample Chord diagram.}
		\label{fig:A sample Chord diagram}
	\end{center}
\end{figure}

The proof proceeds by computing the moments 
\begin{equation}
	m_L={1\over 2^n} E \bigl(\text{Tr}(H^L) \bigr)
\end{equation}
in the following steps:

1) For the first step one needs to define what are {\it chord diagrams}. Consider $L=2n$ dots on a circle - a chord diagram is a pairing of these dots into $n$ pairs. We draw a line connecting each paired dots. i.e., a total of $n$ lines. Denote a specific chord diagram by $\pi$. We then denote by $k(\pi)$ the number of crossings of lines (when we draw the diagram such that each pair of lines intersects at most once).
An example of a chord diagram is shown in Figure \ref{fig:A sample Chord diagram} with $n=8$ and with number of crossings $k=2$.

In the first step one shows that 
\begin{equation}\label{ChordPart}
	m_L=\sum_\pi e^{-\lambda k(\pi)}
\end{equation}
where the sum is over all the chord diagrams. The expression on the RHS is called the {\it chord partition function}, and $q=e^{-\lambda}$ was defined before in \eqref{ParamDef} in terms of the parameters of the spin glass. For example, the contribution to the sum by the chord diagram shown in Figure \ref{fig:A sample Chord diagram} would be $e^{-2 \lambda}$. Chord diagrams were also used in \cite{a:2018kvh} for computing $1/N$ corrections in the SYK model.

In section \ref{subsec:spin glass to chord diagrams} we review this step of the proof in more details since that part of the proof will not change. Furthermore, we will also need to slightly temper with it when computing the 2-pt function.

2) In step 2, one uses the Riordan and Touchard formulae  \cite{riordan1975distribution,flajolet2000analytic} and the results of \cite{Ismail:1987} to show that \eqref{ChordPart} are the moments of the distribution \eqref{ESDist} and further give an explicit formulae for the moments as
\begin{equation}\label{RTChord}
	m_L = {1\over (1-e^{-\lambda})^{L \over 2}}\sum_{j=-{L\over 2}}^{L\over 2} (-1)^j e^{-{j(j-1) \lambda \over 2}} { L \choose {L/2+j}}
\end{equation}

\subsection{Summary of new results}\label{sec:Summary of results}\label{subsec:new results}

In this paper we discuss a new proof for the value of $m_L$ and the energy eigenvalue distribution of the spin glass. We use this to compute the exact two point function for random operators, in the limit $\lambda$ fixed,$\ n\to\infty$. Denoting a new random operator by $M$, it has the form \eqref{RandHam}-\eqref{Hamilt} (with new randomly chosen coefficients, uncorrelated with those of the Hamiltonian as mentioned in  the beginning of section \ref{sec:Motivation and Summary}) but with a new parameter length parameter $p' \propto \sqrt n$.

More precisely we show that 
\begin{multline}\label{2PtReslt}
	2^{-n} E\left[ \Tr\bigl( e^{-{\beta  H \over 2}} M(t) e^{-{\beta  H \over 2}} M(0) \bigr)\right] =\\
	{(q;q)_\infty^2 ({\tilde q}^2;q)_\infty\over (2\pi)^2 }\int_0^\pi d\theta_1 d\theta_2 e^{{2 \cos(\theta_1)(-{\beta\over 2}+it)\over\sqrt{(1-q})}} e^{{2\cos(\theta_2)(-{\beta\over 2}-it)\over\sqrt{(1-q})}}  \\
 \ \times { (e^{2i\theta_1},q)_\infty(e^{-2i\theta_1},q)_\infty(e^{2i\theta_2},q)_\infty  (e^{-2i\theta_2},q)_\infty  
		\over ({\tilde q}e^{i(\theta_1+\theta_2)},{\tilde q}e^{i(-\theta_1+\theta_2)},{\tilde q}e^{i(\theta_1-\theta_2)},{\tilde q}e^{i(-\theta_1-\theta_2)};q)_\infty},\ \ \ 		
\end{multline}
where $ \tilde q \equiv e^{-{4\over 3} {pp'\over n}}$ and $(a,q)_\infty$ is the q-Pochammer symbol (see \eqref{qinf-poch}).

To prove this one evaluates 
\begin{equation}
	m_{k_1k_2}=2^n E\left[ \Tr \bigl( M H^{k_1} M H^{k_2} \bigr) \right].
\end{equation}
We show that the relevant chord diagram which computes this two point function is a chord diagram in which one of the lines is marked, and intersections with this chord are assigned a different weight. An example of a marked chord diagram is given in Figure \ref{fig:A sample marked Chord diagram}. More precisely :
\begin{itemize}
	\item Given 2n+2 points on a circle, two specific points are connected. This is the "marked" chord. The thick line in Figure \ref{fig:A sample marked Chord diagram} connecting the red dots represents the marked chord. 
	\item Between the special points at the ends of the marked chord there are $k_1$ regular points on one side, and $k_2$ regular points on the other side ($k_1+k_2=2n$).
	\item These remaining 2n points are paired. These will be called "regular" chords.
	\item Intersection between regular chords is assigned weight $q$, and the intersection between the regular and marked chords is assigned weight $\tilde q$.
	\item The marked chord partition function is defined as a sum over pairings of the $2n$ regular points, with $k_1$ and $k_2$ fixed and with weights as above, i.e.,
	\begin{equation}
		Z_{k_1k_2}=\sum_{\pi \in \text{\small{marked  chord  diagram}}} q^{k_{\text{regular}}(\pi)} {\tilde q}^{k_{\text{marked}}(\pi)} 
	\end{equation} 
	where $k_{\text{regular}}$ ($k_{\text{marked}}$) is the number of regular-regular (regular-marked) intersections. For example the 1-marked chord diagram in Figure \ref{fig:A sample marked Chord diagram} contributes  $q \tilde{q}$ to the $m_{1,5}$.
	\item Similar to \cite{Erdos} we show that
	\begin{equation}
		m_{k_1k_2}=Z_{k_1k_2}
	\end{equation}
	and evaluate the right hand side to obtain \eqref{2PtReslt}
\end{itemize}
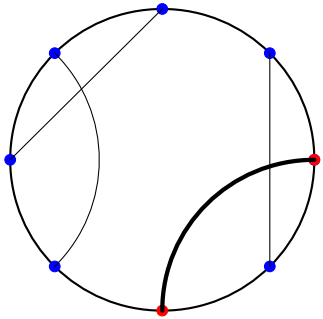
\begin{figure}[t!]
	\def\r{2}
	\begin{center}    
		\begin{tikzpicture}[scale=1,cap=round,>=latex]
		\draw[thick] (0,0) circle(\r);
		\foreach \x in {45,90,135,180,225,315,360} 
		{
			\filldraw[blue] (\x:\r) circle(2pt);
		}
		\foreach \x in {0,270} 
		{
			\filldraw[red] (\x:\r) circle(2pt);
		}
		\draw[ultra thick,color=black] (0,-\r) arc (180:90:\r) ;
		\draw[thin,color=black] (\r/1.414,\r/1.414) -- (\r/1.414,-\r/1.414);
		\draw[thin,color=black] (-\r/1.414,\r/1.414) arc (45:-45:\r) ;        
		\draw[thin,color=black] (-\r,0) -- (0,\r);
		\end{tikzpicture}
		\caption{A sample marked Chord diagram.}
		\label{fig:A sample marked Chord diagram}
	\end{center}
\end{figure}

The evaluation of the various chord partition functions in this work relies on an auxiliary Hilbert space space where there is a natural Hamiltonian whose action is equivalent, in a sense that will be made precise below, to the one of the full Hamiltonian acting on the spin glass Hilbert space. We interpret this auxiliary structure as the bulk dual to the spin glass. Furthermore, we suggest how it reduces to the Schwarzian action in its Liouville form at low energies\footnote{This new Hamiltonian is proportional to $a_q+a_q^\dagger$ where the latter are the creation and annihilation operators of the q-deformed harmonic oscillator.}.


\section{A new derivation of eigenvalue distribution}
\label{sec:new derivation}

Given a random Hamiltonian as in \eqref{RandHam}, the authors in \cite{Erdos} compute the asymptotic distribution of eigenvalues in the $\lambda$-limit \eqref{eq:lambda scaling limit}, by evaluating the moments
\begin{equation}
	m_{L} \equiv \lim_{n \to \infty, \lambda \text{ fixed} } \frac{1}{2^{n}} E \biggl( \Tr  H^L  \biggr)
\label{eq:org-moment}
\end{equation}
and by finding the unique distribution compatible with them. $E\bigl( \bigr)$ stands for an ensemble average. In section \ref{subsec:spin glass to chord diagrams} we review how \cite{Erdos} reduces the moments \eqref{eq:org-moment} to evaluating the chord partition function \eqref{ChordPart}. \cite{Erdos} then uses the results of \cite{Ismail:1987}, and also the formulae of Touchard and Riordan \cite{riordan1975distribution,flajolet2000analytic}, to show that the moments \eqref{ChordPart} arise from the distribution given in \eqref{ESDist}, and to give the explicit formulae \eqref{RTChord} for the moments. Our proof, in section \ref{subsec:Evaluation of the Chord partition function}, replaces this second step, as well as generalizes it to other Chord diagrams, as the one that will appear in the exact 2-pt function in section \ref{sec:2 point function}.

\medskip

\subsection{From spin glasses to chord diagrams}
\label{subsec:spin glass to chord diagrams}

Given the Hamiltonian \eqref{Hamilt}, the computation of the moments 
\begin{equation}
	E\biggl( \Tr(H^L)  \biggr) = \sum_{J_1,..J_L} E\bigl( \alpha_{J_1}.....\alpha_{J_L} \bigr) \Tr\bigl( \sigma_{J_1}...\sigma_{J_L} \bigr)
\end{equation}
proceeds by evaluating the ensemble average of the $\alpha_J$'s first. Any non-vanishing contribution requires at least two insertions of each $\alpha_J$. Moreover, Lemma (4) in \cite{Erdos} shows that the dominant contribution, in the $\lambda$-scaling limit \eqref{eq:lambda scaling limit}, is when $J_1,...J_L$ appear exactly in pairs, with higher multiplicities being subleading in $n$ in the large $n$ limit. This gives us the basic structure of chord diagrams where pairing in the chord is defined by having the same $J$ on two different nodes as in Figure \ref{fig:A sample Chord diagram}. Summing over all the relevant values of $J$ amounts to summing over all chord diagrams (i.e all possible pairings of $J$'s), and then sum over all value of $J$ (i.e., both $\vec e$ and $\vec a$) for each chord. 

Given a chord diagram we therefore need to evaluate what is the weight that is associated with it, i.e.,
\begin{equation}
	\sum_{\text{paired}\ J's} \Tr\bigl( \sigma_{J_1}...\sigma_{J_L} \bigr)
\end{equation}
where there are only $L/2$ independent $J$'s and the pairing is determined by the chord diagram. The obstruction to immediate evaluation is that $\sigma_i^{a}$ for the same site index $i$ can appear in different $J$'s. However, \cite{Erdos} shows that with probability $1$, in the $\lambda$-scaling limit, each node can appear in at most two of the chords, enabling the evaluation of the weight.

More precisely, define the intersection of J's by the intersection of the site index, i.e.
\begin{equation}
	J_i\bigcap J_j=e_i\bigcap e_j.
\end{equation}
\cite{Erdos} shows that, for a given $J_i$ and $J_j$ the size of the overlap is Poisson distributed, and that there is, with probability 1, no triple intersections.  I.e., we can assume
\begin{equation}
	J_i\bigcap J_j\bigcap J_k=0, \ \ i\not= j \not= k \ .
\end{equation}
This statement is summarized in lemma (9) there, and subsequent discussion. Given two sets $a$ and $b$ of integers drawn out of the set $\{1,2,...n\}$ (without repetition in each set), we can think about it as $|a|$ independent processes in which the overlap between the sets increasing by 1 with probability ${|b| \over n}$ (in the limit $n\rightarrow\infty$). This is a Poisson distribution with mean size of overlap ${3 \lambda \over 4} = { |a||b| \over  n}$. Recall that we scale the size of the set with $\sqrt{n}$ so this remains finite in the limit $n\rightarrow\infty$. The average size of an overlap with an additional index set - say $c$ - is the latter times ${ |c| \over n}\rightarrow 0$, so with probability 1, triple overlaps are empty.

The interplay of chord intersections and overlap of the index sets is the key for evaluating the weight of each chord diagram. As we sum over the $J$'s of the chords, then if two chords do not intersect they will contribute  
\begin{equation}\label{NonInter}
	\Tr \bigl (\sigma_{J_1}\sigma_{J_1}\sigma_{J_2}\sigma_{J_2}\bigr)
\end{equation}
whereas if the chords intersect they give a factor proportional to
\begin{equation}\label{InterSig}
	\Tr\bigl (\sigma_{J_1}\sigma_{J_2}\sigma_{J_1}\sigma_{J_2}\bigr)
\end{equation}
If there is a non-trivial overlap $J_1\bigcap J_2\not= 0$, then these factors will be different. So there is some "penalty" that we pay for each intersection.

More precisely, given a chord diagram (i.e. a pairing $\pi$), recall that $k(\pi)$ is the number of pairwise chord intersection. Each chord intersection has a Poisson distributed overlap of sites. Each overlap is independent of the overlap of the other chord intersection. Each overlap of sites (for a given intersection) comes with a factor
\begin{equation}
	3^{-2}\sum_{a,b=1}^3 {1\over 2} \Tr\bigl( \sigma^{(a)}\sigma^{(b)}\sigma^{(a)}\sigma^{(b)} \bigr)= -{1\over 3},
\end{equation}
relative to $1$ when the ordering is $(aabb)$ which originates from an overlap in a pair of non-intersecting chords.
Therefore, the size, $m$,  of each overlap is Poisson distributed with expectation value ${3 \lambda\over 4}$ and comes with a weight $(-{1 \over 3})^m$. The expectation value of the weight for each chord intersection is therefore $e^{-\lambda}$, and the total weight associated with each chord diagram is $e^{-\lambda k(\pi)}$. Hence, one finally obtains \eqref{ChordPart}.

\subsection{Evaluation of the Chord partition function}\label{subsec:Evaluation of the Chord partition function}

In this subsection, we will provide a alternative derivation of the chord partition function reproducing the expression for $v(E|q)$ in \eqref{ESDist}. The proof is rather compact, generalizes to more complicated chord partition functions, such as the ones discussed in section \ref{sec:2 point function}\footnote{Which are not evaluated in the mathematical literature, to the best of our knowledge.} and suggests a bulk interpretation that we develop in section \ref{sec:Bulk}. 
The evaluation of \eqref{ChordPart} is based on a "hopscotch" recursion relation satisfied by the concept of a partial, or open, chord partition function as follows:
\begin{itemize}
	\item $2^{-n}\Tr(H^L)$ involves $L$ points in a chord diagram, as indicated in figure \ref{fig:A sample Chord diagram}. Choose one point, i.e. choose one of the $H$ factors, to be the first and begin moving clockwise in the chord diagram. Each time one reaches an extra point and hop over it, we shall refer to it as "a step". As we go along, denote the number of such steps by $i$, i.e. the number of $H$ factors that were hopped over. In step 1 we hopped over the factor of H that we chose to be the first.
	\item At the $i$'th step, a chord can end on the new $i$'th point, or a new chord can emanate from it. The collection of these decisions defines a chord history. Denote the number of chords that remain opened at this point (open chords) by $l$. This number ranges between 0 and $L$. 
	\item Each open chord is assigned a vertical position relative to the other open chords. Chords emanating from a point further from the left are higher than chords emanating to their right (of course, all chords have emanated left of where we are at right now in the diagram). This book keeping guarantees that open chords have not intersected yet. However, since in previous steps chords have already emanated and ended on various points, our procedure may have taken us through chord configurations of many chords that have already closed to the left of our current position. 
	\item Define $\Pi(i,l)$ as the set of chord histories ending with $l$ open chords at step $i$, and define the partial, or open, chord partition function $v^{(i)}_l$ as
	\begin{equation}
		v^{(i)}_l=\sum_{\pi\in \Pi(i,l)} q^{-k_p(\pi)} \quad \text{with} \quad v^{(0)}_l=\delta_{l,0}.
	\end{equation}
	Here $k_p(\pi)$ refers to the number of chord intersections to the left, in the past of our "hopscotch" process. It is convenient to think about $v^{(i)}$ as a column vector and $l$ as its index. 
\end{itemize}

Given this set-up, one can write down a recursion relation for $v^{i}$. At each step, one can either close a chord (as in Figure \ref{fig:Chord Diagram Recursions : A previous line closes}) or start a new one (as in Figure \ref{fig:Chord Diagram Recursions : A new line begins}) at the point one is hopping over. 
\begin{figure}[t!]
	\def\s{2}
	\begin{center}    
		\begin{tikzpicture}[scale=0.9,cap=round,>=latex]
		\foreach \x in {1,...,4} 
		{
			\filldraw[blue] (\x*\s,0) circle(2pt);
		}
		\foreach \x in {8,10} 
		{ 
			\draw[thin,dashed] (0*\s,\x*\s/10) -- (0.3*\s,\x*\s/10);
			\draw[thin] (0.3*\s,\x*\s/10) -- (3.5*\s,\x*\s/10);
			\draw[thin,dashed] (3.5*\s,\x*\s/10) -- (4.5*\s,\x*\s/10);
		}        
		
		\draw[thin] (1.5*\s,6*\s/10) -- (3.5*\s,6*\s/10);
		\draw[thin,dashed] (3.5*\s,6*\s/10) -- (4.5*\s,6*\s/10);
		
		\draw[thin,dashed] (0*\s,7*\s/10) -- (0.3*\s,7*\s/10);
		\draw[thin] (0.3*\s,7*\s/10) -- (1.3*\s,7*\s/10);
		
		\draw[thin,dashed] (0*\s,9*\s/10) -- (0.3*\s,9*\s/10);
		\draw[thin] (0.3*\s,9*\s/10) -- (2.19*\s,9*\s/10);
		
		\draw[thin,color=black] (2*\s,0) arc (0:90:7*\s/10) ;
		\draw[thin,color=black] (\s,0) arc (180:90:6*\s/10) ;		
		\draw[thin,color=black] (3*\s,0) arc (0:90:9*\s/10) ;		
		
		\draw[thick, ->](2.8*\s,0) to [out=-90,in=-180] (3.05*\s,-0.2*\s);		
		\draw[thick](3.0*\s,-0.2*\s) to [out=0,in=-90] (3.2*\s,0);
		\draw[thick, color=black]
		{
			(3*\s,-0.5) node [below] {\small{Step $i$}}
			(2*\s,10*\s/10) node [above] {\small{ $ \vdots $}}
		};
		\end{tikzpicture}
		\caption{Chord Diagram Recursions : A previous line closes}
		\label{fig:Chord Diagram Recursions : A previous line closes}
		\begin{tikzpicture}[scale=0.9,cap=round,>=latex]
		\foreach \x in {1,...,4} 
		{
			\filldraw[blue] (\x*\s,0) circle(2pt);
		}
		\foreach \x in {8,...,10} 
		{ 
			\draw[thin,dashed] (0*\s,\x*\s/10) -- (0.3*\s,\x*\s/10);
			\draw[thin] (0.3*\s,\x*\s/10) -- (3.5*\s,\x*\s/10);
			\draw[thin,dashed] (3.5*\s,\x*\s/10) -- (4.5*\s,\x*\s/10);
		}        
		
		\draw[thin,dashed] (3.5*\s,5*\s/10) -- (4.5*\s,5*\s/10) ;
		
		\draw[thin,dashed] (3.5*\s,6*\s/10) -- (4.5*\s,6*\s/10) ;
		\draw[thin,color=white] (0*\s,6*\s/10) -- (1.5*\s,6*\s/10) ;		
		\draw[thin] (1.6*\s,6*\s/10) -- (3.5*\s,6*\s/10);

		\draw[thin,color=black] (2*\s,0) arc (0:90:7*\s/10) ;
		\draw[thin,dashed] (0*\s,7*\s/10) -- (0.3*\s,7*\s/10);
		\draw[thin] (0.3*\s,7*\s/10) -- (1.3*\s,7*\s/10);

		\draw[thin,color=black] (\s,0) arc (180:90:6*\s/10) ;		
		\draw[thin,color=black] (3*\s,0) arc (180:90:5*\s/10) ;		
		
		\draw[thick, ->](2.8*\s,0) to [out=-90,in=-180] (3.05*\s,-0.2*\s);		
		\draw[thick](3.0*\s,-0.2*\s) to [out=0,in=-90] (3.2*\s,0);
		\draw[thick, color=black]
		{
			(3*\s,-0.5) node [below] {\small{Step $i$}}
			(2*\s,10*\s/10) node [above] {\small{ $ \vdots $}}
		};
		\end{tikzpicture}
		\caption{Chord Diagram Recursions : A new line begins}
		\label{fig:Chord Diagram Recursions : A new line begins}
	\end{center}
\end{figure}
If one starts a new line, $l$ changes to $l+1$ and the new line enters at the bottom. If one closes a line, it can be either of the $l$ open chords with height between $1$ and $l$. If one closes the line at height $p$, it crosses $(p-1)$ lines on its way down. This crossing generates a weight  $q^{p-1}$ when evaluating its contribution to the partial chord partition function. Altogether, the vector of such partition functions satisfies the following recursion relation
\begin{equation}
	v^{(i+1)}_l = v^{(i)}_{l-1}+(1+q+...+q^{l})v^{(i)}_{l+1}
\label{eq:recursion}
\end{equation}
with initial condition $v^{(0)}_l=\delta_{l,0}$. The latter can be rewritten in terms of an $(L+1)\times (L+1)$ transfer matrix $T_{(L)}$
propagating the partial chord partition function forward
\begin{equation}
	v^{(i+1)}=T_{(L)}v^{(i)}\,,
\label{eq:eigen-problem}
\end{equation}
with matrix elements (indices running from $0$ to $L$, $l_1$ ($l_2$) is the row (column) index)
\begin{equation}
[\ T_{(L)}\ ]^{\ l_2}_{l_1}=\delta_{l_1-1}^{l_2} + \eta_{l_1} \delta_{l_1+1}^{l_2}, \hspace{20mm} \eta_l=1+q+...+q^l= {1-q^{l+1}\over 1-q}
\label{eq:T-matrix}
\end{equation}
describing a matrix with 1's and $\eta_l$'s on the diagonal below and above the main diagonal, respectively, i.e.
\begin{equation}
	\label{eq:T Matrix form}
	T_{(L)} = \begin{bmatrix} 
		0 & {1-q \over 1-q}&  0  &  0 & 0 &  \dots   \\
		1 & 0 &{1-q^2 \over 1-q} &  0 & 0 &  \dots  \\
		0 &  1 & 0 & {1-q^3 \over 1-q} & 0 & \dots \\
		\vdots  &  \ddots & \ddots & \ddots &  \ddots & \ddots \\
	\end{bmatrix}_{(L+1) \times (L+1)}
\end{equation}
To compute the chord partition function \eqref{ChordPart}, define the vector 
\begin{equation}
	|0\rangle_L= v^{(0)} =(1,\!\underbrace{0\dots,0}_{L \text{ entries}})^\intercal
\end{equation}
of length $L+1$, and then 
\begin{equation}
	m_L = {}_L\langle 0| T_{(L)}^L |0\rangle_L\ .
\end{equation}
The initial condition $v^{(0)}_l=\delta_{l,0}$ dictates the use of the initial state  $|0\rangle_L$. Ensuring our procedure counts only chord diagrams that close by the time we reach the $L$-th point, such that we are computing the usual chord partition function in which all lines are paired, determines the final state.  

Notice that we are computing the trace of $H^L$ in the original $2^n$ dimensional Hilbert space, using some auxiliary space based on partial chord diagrams. We shall develop a "bulk" interpretation for the latter in section \ref{sec:Bulk}.
	
Next, given some fixed $L$, one can always consider a larger L'-sized Hilbert space $(L<L')$ such that
\begin{equation}
	m_L =  {}_{L'}\langle 0|  T_{(L')}^{L} |0\rangle_{L'} ,\quad \quad L' \ge L
\end{equation}
This allows us to take $L'\to \infty$, keeping $L$ fixed. In this infinite dimensional Hilbert space, one can define
\begin{equation}
	T \equiv \lim_{{L'}\to \infty} T_{(L')},\hspace{20mm} |0\rangle \equiv (1,0,0,....)^\intercal.
\end{equation}
Hence, $T$ is the infinite dimensional extension of \eqref{eq:T Matrix form}. This provides an auxiliary Hilbert space and a single matrix $T$ in which one can evaluate all traces as
\begin{equation}\label{eq:moments as a trace of T}
	m_L = \langle 0| T^L |0\rangle
\end{equation}

The problem of computing the moments \eqref{ChordPart} reduces to the problem of computing the eigenvalues $\alpha$ of the operator $T$ and expanding the vector $|0\rangle$ in terms of these eigenvectors $|\alpha\rangle$, in an expression
\begin{equation}\label{eq:Eigenvalue decomposisiton}
  m_L= \int_{\text{Spec}(T)} d\alpha \  \alpha^L \  \rho(\alpha) \ |\psi_0(\alpha)|^2\,,
\end{equation}
where $\text{Spec}(T)$ is the set of eigenvalues, $\rho(\alpha)$ is its density and $\psi_0(\alpha) \equiv \langle 0| \alpha \rangle$ is the overlap of $|0\rangle$ with the $|\alpha\rangle$ eigenvector of $T$. Fortunately, $\text{Spec}(T)$ and the density are very easy to compute and the overlap is given by specific q-Hermite polynomials, as we will see below.

In the notation of the spin glass model, comparing the $L$ dependence in the original moment \eqref{eq:org-moment} with $\alpha^L$ in equation \eqref{eq:Eigenvalue decomposisiton}, suggests the identification
\begin{equation}
	\alpha = E
\end{equation}
where $E$ is the energy of the system, properly interpreted. The asymptotic distribution of the energies should then be identified as 
\begin{equation}\label{DensToDens}
	v(E|q) = \rho(E) |\psi_0(E)|^2\,.
\end{equation}

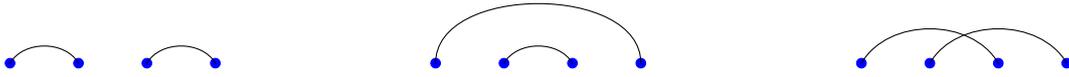
\begin{figure}[t!]
	\begin{center}    
		\begin{tikzpicture}[scale=0.9,cap=round,>=latex]
        \def\c{0} 		
		\foreach \x in {0,1,2,3} 
		{
			\filldraw[blue] (\c+\x,0) circle(2pt);
		}
		\draw[thin,color=black] (\c + 0,0) to [out=60,in=180-60] (\c+1,0);
		\draw[thin,color=black] (\c + 2,0) to [out=60,in=180-60] (\c+3,0);
		 \end{tikzpicture}
 \hspace{25mm} 		
		\begin{tikzpicture}[scale=0.9,cap=round,>=latex]
		      \def\c{0}
		\foreach \x in {0,1,2,3} 
		{
			\filldraw[blue] (\c+\x,0) circle(2pt);
		}
		\draw[thin,color=black] (\c + 0,0) to [out=90,in=180-90] (\c+3,0);
		\draw[thin,color=black] (\c + 1,0) to [out=60,in=180-60] (\c+2,0);
        \end{tikzpicture}
\hspace{25mm}
		\begin{tikzpicture}[scale=0.9,cap=round,>=latex]
		\def\c{10} 
		
		\foreach \x in {0,1,2,3} 
		{
			\filldraw[blue] (\c+\x,0) circle(2pt);
		}
		\draw[thin,color=black] (\c + 0,0) to [out=60,in=180-60] (\c+2,0);
		\draw[thin,color=black] (\c + 1,0) to [out=60,in=180-60] (\c+3,0);
		\end{tikzpicture}
	\end{center}
\caption{Chord Diagram for $L=4$ case}
\label{fig:neq4 Chord Diagrams}\end{figure}

\paragraph{A short example.}
It is worth while carrying out the procedure above in an explicit, low $L$ case, and compare the result with \eqref{RTChord}. For example $m_4(q) = 2 + q$, which can obtained also from the three chord diagrams in 
 Figure \ref{fig:neq4 Chord Diagrams}. In our approach we start with $v^{(0)}$, act on it 4 times with $T_{(4)}$ (or $T$), and project on $v^{(0)}$. Keeping track of chord histories give the following partial chord partition functions: 
 \begin{equation}
 v^{(0)} = \begin{pmatrix}
 1 \\
 0 \\
 0 \\
 0 \\
 0 \\
 \end{pmatrix} \hspace{5mm}  v^{(1)} = \begin{pmatrix}
 0 \\
 1 \\
 0 \\
 0 \\
 0 \\
 \end{pmatrix} \hspace{5mm}  v^{(2)} = \begin{pmatrix}
 1 \\
 0 \\
 1 \\
 0 \\
 0 \\
 \end{pmatrix} \hspace{5mm}  v^{(3)} = \begin{pmatrix}
 0 \\
 2+q \\
 0 \\
 1 \\
 0 \\
 \end{pmatrix} \hspace{5mm} v^{(4)} = \begin{pmatrix}
 2+q \\ 0 \\ 3+2 q+q^2 \\ 0 \\ 1 \\
 \end{pmatrix}
 \end{equation} 
 
\paragraph{The symmetric form of the transfer matrix T.} The matrix $T$ in \eqref{eq:T Matrix form} is not Hermitian, but one can conjugate it to a symmetric version by defining a new matrix ${\hat T}$
\begin{equation}
	\hat T \equiv P T P^{-1} 
\end{equation}
where $P$ is a diagonal matrix with entries $(P_0,P_1,P_2\dots )$ satisfying
\begin{equation}
	P_l = \prod_{i=0}^{l-1} \sqrt{\eta_i} ={ \sqrt{ (q;q)_l } \over (1-q)^{l \over 2}}\,, \quad l\neq 0 \quad \quad P_0=1\,,
\end{equation}
where $(a;q)_{l}$ is the q-Pochammer symbol (see \eqref{q-poch}). $\hat T$ has matrix elements
\begin{equation}
  (\hat T)^{\ l_2}_{l_1} = \sqrt{\eta_{l_2}}\delta^{l_2}_{l_1-1} + {\sqrt{\eta_{l_1}}} \delta^{l_2}_{l_1+1}.
\label{eq:sym-T}
\end{equation}
Thus, it is manifestly symmetric,
\begin{equation}\label{eq:The Sym Transfer Matrix}
	\hat T = \begin{bmatrix} 
		0 & 1 &  0  &  0 & 0 &  \dots   \\
		1 & 0 & \sqrt{\eta_1}&  0 & 0 &  \dots  \\
		0 &  \sqrt{\eta_1} & 0 & \sqrt{\eta_2}& 0 & \dots \\
		\vdots  &  \ddots & \ddots & \ddots &  \ddots & \ddots \\
	\end{bmatrix}
\end{equation}
and has the same original moments \eqref{eq:moments as a trace of T} since 
\begin{equation}
	m_L= \langle 0 | T^L | 0 \rangle  = \langle 0 | P^{-1}  \hat T^L P | 0 \rangle  = \langle 0 |   \hat T^L   | 0 \rangle\,. 
\end{equation}
We will switch between the two transfer matrix descriptions depending on which is more convenient at each stage.

\subsubsection{The spectrum of T}\label{subsubsec:Spectrum of T}

Obtaining the spectrum of T is straightforward. The matrix T asymptotes, down the diagonal, to a matrix with 1 one diagonal below the main diagonal and $\eta_\infty={ 1 \over 1-q}$ one diagonal above the main diagonal. i.e.
\begin{equation}\label{eq:T asymp}
	T_{\text{asymp}} \equiv \begin{bmatrix} 
		0 & {1 \over 1-q}&  0  &  0 & 0 &  \dots   \\
		1 & 0 &{1 \over 1-q} &  0 & 0 &  \dots  \\
		0 &  1 & 0 & {1 \over 1-q}  & 0 & \dots \\
		\vdots  &  \ddots & \ddots & \ddots &  \ddots & \ddots \\
	\end{bmatrix},\ \ \  \delta T \equiv T - T_\text{asymp} 
\end{equation}
We can think about the eigenvalue problem of T 
\begin{equation}
	T\psi=E \psi
\end{equation}
as a scattering problem with the distance along the diagonal playing the role of position. In this interpretation, infinity is captured by the asymptotic form of the operator $T$ far down the diagonal. Hence, this is a scattering problem on the half line with $\delta T$ acting as a scatterer close to the origin. Indeed, up to an overall rescaling, by conjugating the matrix $T_{asymp}$, and adding the identity matrix with an appropriate weight, we can bring it to the form with -2 on the diagonal and 1 on the diagonals below and above the main. It is then an approximation to the 2nd derivative operator, making the asymptotic behaviour more familiar in the continuum limit. This interpretation is elaborated in section \ref{sec:Bulk}, where the connection between this eigenvalue problem and the Liouville equation is described. 

However, as far as the spectrum and its density is concerned, the details of the scatterer are not important as both can be read from the behaviour at infinity\footnote{For the wave functions, or form factors, we will be more specific below. Also, reading the spectrum and density from infinity also assumes that there are no bound states near the origin.}. So the spectrum of $T$ is the same as that of $T_{\text{asymp}}$ which is a Toeplitz tridiagonal matrix (i.e with constant elements one diagonal above and below the main diagonal \cite{NLA:NLA1811}), for which there is a simple formula for the eigenvalues, which in this case is
\begin{equation}
	-\frac{2}{\sqrt{1-q} }\cdot \cos{s \pi \over  L'+1},\ s=1,\dots L'\ \ \ \xrightarrow{L'\to\infty} \bigl( \frac{-2}{\sqrt{1-q}}, \frac{2}{\sqrt{1-q}} \bigr)
\end{equation}
as was found in \cite{Erdos}. This formula gives us both the spectrum of the $T$ and the density of states on it. Denote $\theta={ s \pi \over n+1}$	, then in the limit $n\rightarrow \infty$ it covers the interval $[0,\pi]$ with uniform distribution, i.e., inserting a complete set of energy eigenstates is simply done by the replacement
\begin{equation}
	\sum_E \to \int_0^\pi d\theta
\end{equation}

\subsubsection{Eigensystem of $T$ matrix}
\label{eq:eigensystem}

The previous asymptotic discussion suggests to parametrise the eigenvalues of the matrix $T$ as $E(\mu) \equiv {2 \mu \over \sqrt{1-q} }$. Let $v^{(\mu)}$ be the corresponding eigenvector\footnote{$v^{(\mu)} \propto \psi(\alpha)$ of section \ref{sec:new derivation}. For now the normalization is different though.}. This allows us to write  \eqref{eq:eigen-problem} together with the recursion relation \eqref{eq:recursion} as
\begin{equation}
T \cdot v^{(\mu)} = {2\mu \over \sqrt{1-q} }  v^{(\mu)}\   \rightarrow\ {2\mu\over \sqrt{1-q}} v^{(\mu)}_l =   v^{(\mu)}_{l-1}  + { (1-q^{l+1}) \over (1-q) } v^{(\mu)}_{l+1}\,, \quad v_0^{(\mu)}=1
\label{eq:eigen-1}
\end{equation}
Just for this subsection, we will allow $l=-1$ and define $v_{-1}^{(\mu)}=0$. The recursion relation is simplified by working with the new vectors $u_l^{(\mu)}$
\begin{equation}
	v_l^{(\mu)} = { (1-q)^{l \over 2}    \over (q;q)_{l}   } u_l^{(\mu)}
\end{equation}
so that \eqref{eq:eigen-1} becomes 
\begin{equation}
	2\mu   u_l^{(\mu)} = (1-q^l) u_{l-1}^{(\mu)} + u_{l+1}^{(\mu)} \hspace{20mm} u_{-1}^{(\mu)}=0, u_0^{(\mu)}=1
\end{equation}
Comparing with the recursion relations satisfied by continuous q-Hermite polynomials given in \eqref{eq:Recursion of H_n}, we can identify $u_l^{(\mu)}$ with a q-Hermite polynomial $H_l(\mu|q)$ and $\mu$ with $\cos \theta$, with $\theta \in [0,\pi]$. Hence, the eigenvector of the transfer matrix $T$ equals 
\begin{equation}\label{eq:Eigenvector of T and its Range}
	v_l^{(\mu)} =  { (1-q)^{l \over 2}    \over (q;q)_{l}   } H_l(\mu|q) \hspace{30mm} -1 \le \mu \equiv \cos(\theta) \le 1 
\end{equation}
Using the full range $\theta$ is dictated by the the discussion of the spectrum in section \ref{subsubsec:Spectrum of T}.

At this point, it is more useful to switch to $\hat T$, since we need to conjugate the form factor. The eigenvectors of symmetric transfer matrix $\hat T$ are just $P v^{\mu}$. In components, the eigenvectors are 
\begin{equation}\label{TEigV}
\hat \psi_l(\theta) =  N(\mu,q)  \ P_l v^{(\mu)}_l =N(\mu,q) \ { H_l( \mu |  q ) \over \sqrt{ (q;q)_l } }.  \hspace{20mm} \mu = \cos \theta
\end{equation}
where $N(\mu,q)$ is a normalization which is fixed by the requirement that the states are delta function normalized in $\theta$, which gives (see Appendix \ref{section Appendix:Some properties of Continuous q-hermite polynomials}) $N(\mu,q) = {\sqrt{(q ;q)_\infty} |(   e^{2i \theta};q)_\infty| \over \sqrt{2\pi} }$. With this one can easily write down the matrix element \footnote{Recall that the density of states $\rho(\theta)$ is uniform.}
\begin{equation}\label{eq:hat T matrix element}
\langle l | \hat T^L | m \rangle  = \int_0^\pi d\theta \ \hat \psi_l(\theta)  \hat \psi_m(\theta) E(\theta)^L
\end{equation}
The moments of the distribution eq(\ref{eq:Eigenvalue decomposisiton}) can then be computed to be
\begin{equation}\label{eq:Our moment formuale}
	m_L(q) = \int_0^\pi d\theta   { (q;q)_\infty  |(e^{2 i \theta} ; q)_\infty|^2 \over  2\pi}  E(\theta)^L = \int_0^\pi d\theta \ \Psi(\theta,q)
	E(\theta)^L
\end{equation}
where we have defined the distribution
\begin{equation}\label{eq:FormFacA}
 \Psi(\theta,q) \equiv |\hat \psi_0(\theta)|^2 = { (q;q)_\infty  |(e^{2 i \theta} ; q)_\infty|^2 \over  2\pi}
\end{equation}
Below we will show that this formula is the same distribution given in \eqref{ESDist}.

\paragraph{Matching to the result in \cite{Erdos}.}
Recall that \cite{Erdos} obtained the moments $m_L(q)$ as moments of distribution $v(E|q)$ given in \eqref{ESDist}, i.e 
\begin{equation}\label{eq:Erdos Moments formualae in E}
	m_L = \int_{-{2\over \sqrt{1-q} }}^{2\over \sqrt{1-q} } dE\ v(E|q) E^k  
\end{equation}
Switching to angular variables via $E = E(\theta)\equiv {2 \cos\theta \over \sqrt{1-q} }$, we write $v(E|q)$ as
\begin{equation}\label{eq:Erdos form factor simplified}\begin{split}
		v(E(\theta) | q)  
		& =  {\sqrt{1-q} \over \pi \sin \theta} \prod_{k=0}^\infty {(1-q^{2k+2}) \over (1-q^{2k+1})(1+q^k)^2}  \left\lbrace (1 - e^{2i \theta} q^{k})(1 - e^{-2i \theta} q^{k} ) \right\rbrace \\
		& = {\sqrt{1-q} \over 4\pi \sin \theta} (q;q)_\infty |(e^{2i\theta};q)_\infty|^2
	\end{split}
\end{equation}
followed by a change of variables as
\begin{equation}
	d\theta \ \Psi(\theta,q) = dE  \ v(E(\theta) | q)
\end{equation}
to obtain that \eqref{eq:Erdos Moments formualae in E} matches our result \eqref{eq:Our moment formuale}.


\section{The $q\to 1$ limit of the distribution}
\label{sec:qeq1 limit}

Our interest in approximately factorized correlators, suggests to work in the regime $\lambda\to 0$ or $q\to 1$. The analysis is most easily done by arranging the Pochhammer symbols into Jacobi Theta functions and performing modular transformations. The results in this section are similar to Appendix B in \cite{Cotler:2016fpe} and \cite{Garcia-Garcia:2017pzl} after a suitable substitution that takes us between our model and the SYK model in the same scaling as above. 

To study the $\lambda\to 0$ limit of the distribution $\Psi(\theta,q)$ \eqref{eq:FormFacA}, it is convenient to rewrite it in terms of  Jacobi Theta functions. Using \eqref{Theta-poch}, 
\begin{equation}
  \Psi(\theta,q) =  { \sin \theta \  \vartheta_1({\theta \over \pi} | {i \lambda \over 2\pi})   \over  \pi q^{1 \over 8}}
\end{equation}
Using the modular transformation \eqref{eq:Modular Transformation Formulae} we rewrite it as 
\begin{equation}
\vartheta_1({ \theta \over \pi } |{i \lambda \over 2\pi} ) = \vartheta_1({2i \theta \over \lambda} | {2 \pi i \over \lambda} ) e^{- {2   \theta^2 \over \lambda}} {1 \over i} \sqrt{ 2\pi \over \lambda}\,.
\label{eq:mod-1}
\end{equation}
and the $\lambda \to 0$ limit becomes
\begin{equation}
\begin{split}
\vartheta_1({2i \theta \over \lambda} | {2 \pi i \over \lambda} )
& = 2 e^{- {\pi^2 \over 2 \lambda}} \sin({2\pi i \theta \over \lambda}) \prod_{m=1}^\infty (1 - e^{-{4 m \pi^2 \over  \lambda}}) (1 - 2 \cos({4\pi i \theta \over \lambda})e^{-{4 m \pi^2 \over  \lambda}} + e^{-{8 m \pi^2 \over  \lambda}}  ) \\
& \xrightarrow{\lambda \to 0} 2 i e^{- {\pi^2 \over  2\lambda}} \sinh({2\pi \theta \over \lambda}) \prod_{m=1}^\infty \left( 1 -2 \cosh ({4 \pi \theta \over \lambda}) e^{-4 \pi^2 m \over \lambda} \right) \\
& \hspace{5mm}=  2 i e^{- {\pi^2  \over 2\lambda}} \sinh({2\pi \theta \over \lambda})  \left( 1 -2 \cosh ({4 \pi \theta \over \lambda}) e^{-4 \pi^2  \over \lambda} \right)\ .
\end{split}
\end{equation}
The last equality follows since the exponential overcomes the hyperbolic cosine factor for $m \ge 1$ given that $\theta \le \pi$. Plugging the above $\lambda \to 0$ expansion in \eqref{eq:mod-1}, one gets 
\begin{equation}\label{eq:qeq1 of Jacobi Theta}
\begin{split}
\vartheta_1({ \theta \over \pi } |{i \lambda \over 2\pi} ) & =2 \sqrt{2 \pi \over \lambda}  e^{-{2 \theta^2 \over \lambda}-{\pi^2 \over 2 \lambda}} \sinh({2\pi \theta \over \lambda})  \left( 1 -2 \cosh ({4 \pi \theta \over \lambda}) e^{-4 \pi^2  \over \lambda} \right) \\
& \approx 4 \sqrt{2 \pi \over \lambda} e^{- {2\pi^2 \over \lambda}} \ \ 
e^{-{2 \over \lambda} (\theta - {\pi \over 2})^2   } 
  \sinh({2 \pi \theta \over \lambda}) \sinh({2 \pi (\pi - \theta) \over \lambda})
\end{split}
\end{equation} 
where in the last step we used $\theta\ge 0$.  This determines the dominant contribution to the distribution \eqref{eq:FormFacA} to be 
\begin{equation}\label{eq:rho formulae}
\Psi(\theta,q) \approx 4 \sqrt{2 \over \pi \lambda} e^{- {2\pi^2 \over \lambda}} \,
e^{-{2 \over \lambda} (\theta - {\pi \over 2})^2   } 
\sin(\theta)  \sinh(\frac{2 \pi \theta}{\lambda}) \sinh({2 \pi (\pi - \theta) \over \lambda})
\end{equation}
Notice this function is symmetric under $E \to -E$, and vanishes at the edges $E_{max}=-E_{min}={2 \over\sqrt{\lambda}}$, which correspond to $\theta=0$ and $\theta=\pi$, respectively, since $E = \frac{2 \cos(\theta)}{\sqrt{1-q}}$. 

The distribution \eqref{eq:rho formulae} has several interesting regimes:
\begin{itemize}

\item The $\lambda\to 0$ with $E$ fixed regime. As highlighted in \cite{Erdos}, pointwise,
\begin{equation}
  \Psi(\theta,q)  \propto e^{-{E^2\over 2}}\,, \,\,\,\, E={-2\over \sqrt{\lambda}} \bigl(\theta-{\pi\over2}\bigr)\,,
\end{equation}
which is the Gaussian limit of the distribution \eqref{eq:rho formulae}.

\item The other interesting behaviour is close to edges. Setting $\varphi=\pi-\theta$, we begin with the limit
\begin{equation}
\varphi=\pi-\theta \propto \lambda 
\end{equation}
(where by this we also include ${ \varphi \over \lambda} \gg 1 \ \text{fixed},\  \lambda\rightarrow 0$).
In this case the distribution becomes
\begin{equation}
\begin{aligned}
  \Psi(\theta,q) 
  &= 2 \sqrt{2 \over \pi \lambda}  
  e^{-{2 \over \lambda} ( {\pi \over 2} - \varphi)^2  - {2 \pi \varphi \over \lambda}} 
  \sin(\varphi)  \sinh({2 \pi \varphi \over \lambda}) \\
   &= 2 \sqrt{2 \over \pi \lambda}  
  e^{-{\pi^2 \over 2 \lambda} -{2 \varphi^2 \over \lambda} } 
  \sin(\varphi)  \sinh({2 \pi \varphi \over \lambda})   \equiv \Psi(\varphi;q)\,,
\end{aligned}
\label{eq:Def psi varphi}
\end{equation}
The quadratic term in the exponential can also be neglected in this regime, giving rise to the density of states of the Schwarzian theory ($\propto \sinh(2\pi\sqrt{(E-E_\text{min}) / \lambda^{3 \over 2}})$) after recalling that near the edge 
\begin{equation}
E-E_\text{min}={\varphi^2 \over \sqrt{\lambda}}
\end{equation}

\item Note that we can actually neglect the quadratic term in the exponential already at $\lambda \ll \varphi \ll \sqrt{\lambda}$, and extend the Schwarzian regime. This points to another simplification of the spectrum which actually covers the bulk of the spectrum at $\lambda \ll \varphi \ll \pi - \lambda$. In this range we can also expand the second sinh, and obtain that the distribution is just a gaussian in $(\varphi-\pi/2)$. The center of this range includes the Gaussian-in-energy distribution, and its edges overlaps with the Schwarzian distribution. It would be interesting to find a symmetry argument for this entire range.

\end{itemize}



\subsection{The canonical ensemble in the $q \to 1$ limit} 
\label{sec:thermo}

Similarly one can analyze the canonical partition function in the limit $q\rightarrow 1$.
Using the variable $\varphi \equiv \pi - \theta$ as before 
\begin{equation}
  Z(\beta) = \int_0^\pi d\varphi \,e^{2\beta \cos \varphi \over \sqrt \lambda}  \Psi(\varphi;q)\ .
\label{eq:gpart}
\end{equation}

We can treat most of the spectrum using the discussion in bullet 3 above, leaving out only a very low temperature regime where $\phi\sim\lambda$. We will prefer however to split the discussion according to first two bullets, i.e., to a high temperature phase and a Schwarzian phase which then splits into a low temperature and a very low temperature phase. Both of the latter are obtained from the Schwarzian density of states and go smoothly into each other, and we make this division mainly for the sake of the discussion of the 2-pt function in section \ref{sec:2 point function}, for which the difference between these regimes is more meaningful.
 
\begin{itemize}
\item \textbf{High Temperature phase} (when $\lambda^{-{1 \over 2}} \gg \beta$) : localizing in the region  $|\varphi -{\pi \over 2}| \ll 1$, reduces the partition function to a Gaussian around $E=0$, 
and the partition function can then be written as
\begin{equation}
Z(\beta) = \sqrt{2 \over \pi \lambda} \int_0^\pi d\varphi e^{2\beta \cos \varphi \over \sqrt \lambda}  e^{-{2 \over \lambda} ({\pi \over 2} - \varphi)^2   } 
\sin \varphi 
\end{equation} 
It is clear that the gaussian cuts off the integral if $\varphi$ deviates from $\pi \over 2$. To evaluate $Z(\beta)$, we set ${\pi \over 2}-\varphi\equiv x$. The limit above translates to $x \ll 1$, in which case we approximate the cosine by $x^2$ to obtain $Z(\beta)\sim e^{\beta^2 \over 2} $ and
the integral is supported at $x ={\beta\sqrt \lambda \over 2}$ with a width of order $\sqrt \lambda$. 

\item \textbf{Low Temperature phase} (when $\lambda^{-{3 \over 2}} \gg \beta \gg  \lambda^{-{1 \over 2}}$) : generically, one expects $\varphi \ll 1$, but the thermodynamic behaviour of the system is sensitive to how small $\varphi$ is compared to $\lambda$, due to the argument in the sinh factor in \eqref{eq:Def psi varphi}. Consider the regime $\lambda \ll \varphi \ll 1$, where the distribution is approximated by
\begin{equation}
\Psi(\varphi;q) \approx  \sqrt{2 \over \pi \lambda} \, e^{-{2 \over \lambda} ({\pi \over 2} - \varphi)^2 } \varphi   
\end{equation}
Expanding the Boltzmann factor, the partition function reduces to\footnote{The term $-{2 \varphi^2 \over \lambda}$ in the exponent is negligible compared to $-\beta \varphi^2 \over \sqrt \lambda $ due to $\beta \sqrt \lambda \gg 1$.}
\begin{equation}
\begin{aligned}
  Z(\beta) &= \sqrt{2 \over \pi \lambda} \int_0^\pi d\varphi \,e^{2\beta \cos \varphi \over \sqrt \lambda}  e^{-{2 \over \lambda} ({\pi \over 2} - \varphi)^2   } \varphi \\
  &\approx	\sqrt{2 \over \pi \lambda} e^{{2 \beta \over \sqrt \lambda} - {\pi^2 \over 2 \lambda}}\int_0^\pi d\varphi  \,e^{- {\beta \varphi^2 \over \sqrt \lambda} +{2\pi \varphi \over \lambda }} \varphi \approx  {\sqrt 2 \pi \over \beta^{3 \over 2} \lambda^{3 \over 4}} e^{{2 \beta \over \sqrt \lambda} - {\pi^2 \over 2 \lambda}+{\pi^2 \over \beta \lambda^{3 \over 2}}} 
\end{aligned}
\end{equation}
Notice the integral is mainly supported near $\varphi = { \pi \over \beta \sqrt \lambda}$ with a width of $\lambda^{1 \over 4} \over \sqrt \beta$. The consistency with the assumption $\lambda \ll \varphi \ll 1$ requires $\beta  \ll \lambda^{-{3 \over 2}}$. This regime, along with the next one, are part of the conformal low energy limit of the theory.

\item \textbf{Very Low Temperature phase} (when $\beta \gg \lambda^{-{3 \over 2}} $) : consider the regime $\varphi \ll \lambda$. After linearising both the $\sin$ and $\sinh$ factors, the distribution \eqref{eq:Def psi varphi} simplifies to 
\begin{equation}
\Psi(\varphi;q) \approx  {4 \over \lambda} \sqrt{2 \pi \over  \lambda} \ e^{ -{\pi^2 \over 2 \lambda} - {2 \varphi^2 \over \lambda} } \varphi^2 
\end{equation}
The Boltzmann factor in the partition function cuts off the integral around $\varphi \sim {\lambda^{1 \over 4}  \over \sqrt \beta}$. This is consistent with our regime $\varphi \ll \lambda$, since $\beta \gg \lambda^{-{3 \over 2}}$. The partition function can then be evaluated as
\begin{equation}
Z(\beta) = {4 \over \lambda} \sqrt{2 \pi \over  \lambda} e^{{2 \beta \over \sqrt \lambda} - {\pi^2 \over 2 \lambda}}\int_0^\pi d\varphi \,e^{- {\beta \varphi^2 \over \sqrt \lambda}} \varphi^2 \approx  \frac{\pi \sqrt 2}{\beta^{3 \over 2} \lambda^{3 \over 4}}\,  e^{{2 \beta \over \sqrt \lambda} - {\pi^2 \over 2 \lambda}}
\end{equation}
where in evaluating the integral, we have replaced the upper limit by $\infty$. 	

\end{itemize}

\paragraph{Relation to previous work.} The low energy behaviour identified in \eqref{eq:Def psi varphi} is the one discussed in appendix B in \cite{Cotler:2016fpe} and in \cite{Garcia-Garcia:2017pzl}. To make the comparison with \cite{Cotler:2016fpe} easier, notice the density of states \eqref{eq:Erdos form factor simplified} can be written as
\begin{equation}
v(E(\mu) |q) = {\cal N} \ {1 \over \sqrt{1 - \mu^2}} \prod_{k=0}^\infty \left( 1 - {\mu^2 \over \cosh^2({k \lambda \over 2})} \right)
\end{equation}
where $\mu = \cos \theta$ and ${\cal N} = {\sqrt{1-q} \over \pi} \prod_{k=0}^\infty {1 -q^{2 k +2} \over 1 - q^{2k +1}}$. This matches 
equation (81) in \cite{Cotler:2016fpe} by identifying their parameters $a, \lambda_s, {\cal J}$ as
\begin{equation}
  a=\mu\,, \quad {\cal J}=\sqrt{\lambda}\,e^{\lambda/8}\,, \quad \lambda_s= \frac{\lambda}{2}\, 
\end{equation}
whereas both energies are the same\footnote{Our normalizations are different from \cite{Cotler:2016fpe} since their distribution $\rho_s(E)$ integrates to $2^{N \over 2}$ whereas our $v(E|q)$ integrates to $1$}.  The second matching follows from the observation that our variances equal unity, as in \cite{Erdos}, whereas our normalisation was $\text{Tr} H^2/\text{Tr}\mathbb{I} = 1$ (see equation (80) in \cite{Cotler:2016fpe}). The third matching is due to the Majorana nature of the fermions in the SYK model. 

Thus the density of states in \cite{Erdos} is exactly the same as the doubled scaled SYK, up to these identifications. The further triple scaled limit,
\begin{equation}
\lambda \to 0 \hspace{10mm} {2(E-E_\text{min}) \over  \lambda^{3 \over 2}} \mbox{ fixed}\,,
\end{equation}
isolating the Schwarzian action in SYK, corresponds to the low energy behaviour captured by the density of states \eqref{eq:Def psi varphi} in our set-up.

\section{Bulk reconstruction}
\label{sec:Bulk}

In section \ref{sec:new derivation} we presented a new derivation of the density of energies in $v(E|q)$ in the $\lambda$-scaling limit \eqref{eq:lambda scaling limit} 
\begin{equation}\label{Sec4PrtnA}
2^{-n} E\biggl(\Tr(e^{-\beta H})\biggr) = \int_{E_{min}}^{E_{max}} dE v(E|q) e^{-\beta E}
\end{equation}
keeping $\beta$ finite and where $E()$ on the left hand side is the average over the ensemble of Hamiltonians. The range of integration on the right hand side is the spectrum of the random Hamiltonian, and its "randomness" now hides in the $1/n$ corrections which are neglected in this limit. Loosely, one can hope that for a specific realization of the Hamiltonian H, one can write
\begin{equation}\label{Sec4PrtnB}
\Tr(e^{-\beta H}) = \int_{E_{min}}^{E_{max}} dE v(E|q) e^{-\beta E} \bigl(1+{\cal O}({1\over n})\bigr)\ ,
\end{equation}
with probability 1 (or $1-{\cal O}(1/n)$) on the space of random Hamiltonians, in the large n limit. In this case one is dealing with a specific Hamiltonian on the left hand side. This single Hamiltonian realization corresponds to the boundary field theory Hamiltonian in the AdS/CFT written in terms of the fundamental field theory objects - in our case the spin operators. The operator is random and only in the $n\to\infty$ limit its spectrum converges to anything universal.

In this section, we suggest that {\it the operator $T$ (or ${\hat T}$) is the bulk Hamiltonian}, i.e. the analogue of the bulk Hamiltonian for the near-AdS background - whose low energy limit is given by the Schwarzian action - but extended to the full model. Recall that the parameter $E$ appearing in the right hand side of \eqref{Sec4PrtnA} and \eqref{Sec4PrtnB} can be reinterpreted as the energy of the field theory Hamiltonian, but it is also the eigenvalue of the operator $T$ (or ${\hat T}$) which acts on the (altogether different) Hilbert space of weights of open chord lines. Whereas the spectrum of $H$ changes from realization to realization, the matrix ${\hat T}$ is fixed. There is no contradiction since we work in the limit $n\to\infty$, fixed $\lambda$ limit, where the spectrum of $H$ is universal. 

Furthermore, the operator ${\hat T}$ can be used as the Hamiltonian not only for the partition function, but for a much broader set of computations. It should be clear that the insertion of any finite polynomial of $H$ in expectation values involving density matrices of the form
\begin{equation}
\sum_{eigenvalues\ E} |E\rangle f(E) \langle E|
\label{eq:den-mat}
\end{equation}
for any analytic weight function $f(E)$, can be turned into the insertion of the same polynomial with ${\hat T}$ as its argument, following the procedure described in section 
\ref{sec:new derivation}. In other words, the insertion of $e^{-itH}$ in expectation values involving \eqref{eq:den-mat} can be exactly replaced by $e^{-it\hat T}$, while the density matrix itself is mapped into the density matrix (as an operator in the Hilbert space defined on the chord diagram side)
\begin{equation}
f(\hat T)|v_0\rangle  \langle v_0|.
\end{equation}
Having two different Hamiltonians, acting on different Hilbert spaces but propagating the system in exactly the same way, supports the dual interpretation we suggest for ${\hat T}$.

This means that we can access a large set of weights on the energy eigenstates as long as the function is smoother than the energy spacing (actually smoother than ${1 \over n}$ for the entire energy band). This is not in contradiction with what we know about the bulk Hamiltonian (anything which extends the low energy effective action), since it is not clear that it should be able to capture states whose support on close by energy states is rapidly varying\footnote{Unless, for example, one believes in the microstate program in its strongest form where one can choose a specific energy eigenstate in the most extreme case.}.

Phrased differently we regard $E$, when used as the eigenvalue of ${\hat T}$, as a parameter which scans over the allowed energy range only after taking the limit $n\to \infty$. It is not the discrete spectrum of energies of the finite $n$ system. It should be viewed as a coarse grained version of the latter, very much like the energy measured in gravity is a coarse grained version of the discrete set of energies of the field theory (when defined on a compact space). Going from the eigenvalues of $\hat T$ to the eigenvalues of $H$ at finite $n$ is an interesting problem, and it is similar to seeing - in General Relativity - the discreteness in energies of a black hole. 

The above discussion, together with the behaviour of the partition function in the low temperature regime, suggests the low energy physics for $q\to 1$ should be governed by the Schwarzian action (in the gravity dual), as in the SYK model. In the following, we derive this connection by matching the continuum limit of the equation determining the spectrum of $\hat{T}$ with Liouville quantum mechanics\footnote{We would like to thank D. Bagrets for a discussion of this point.}, which can be written as the Schwarzian action, as discussed in \cite{Bagrets:2016cdf,Bagrets:2017pwq}\footnote{See \cite{Mertens:2017mtv} for a 2d CFT perspective on this matter.}.


To take the continuum limit, it is convenient to define the matrix $\tilde T \equiv S \hat T S^{-1}$ where $S$ is a diagonal matrix with entries $S_{ii} = (-1)^i$. Notice that solving for the eigenvalues of the $\tilde T$ matrix still resembles a scattering problem on the half line, with the index $i$ of the vector measuring the distance from the origin, just like it did for the $\hat T, T$ matrices. The asymptotic form of the $\tilde T$ matrix is 
\begin{equation}
\tilde T  =  \frac{1}{\sqrt{1-q}} \begin{bmatrix} 
		0 & -1&  0  &  0 & 0 &  \dots   \\
		-1 & 0 & -1 &  0 & 0 &  \dots  \\
		0 &  -1 & 0 & -1  & 0 & \dots \\
		\vdots  &  \ddots & \ddots & \ddots &  \ddots & \ddots \\
	\end{bmatrix}
\end{equation} 
In the continuum limit, the above matrix includes the second derivative operator. To make this more precise, define
\begin{equation}
\phi=\log(q) i
\end{equation}
Using the form of the $\hat{T}$ operator in \eqref{eq:sym-T}, its continuum limit equals 
\begin{equation}
{\tilde T}\to \frac{1}{\sqrt{1-q}} \biggl( -2 - (\log q)^2\partial_\phi^2 + {q \over 2} e^\phi \biggr)  
\label{eq:cont-T}
\end{equation}
Notice the potential term comes from the expansion $\sqrt{ 1-q^{i+1} \over 1-q } = {1 \over \sqrt{1-q} } (1 -  {q^{i+1} \over 2} +\dots )$ in $\eta_i$, as defined in \eqref{eq:T-matrix}, which is accurate since $i$ is large and $q\to 1$, from below.

The eigenvalue problem then reduces to the quantum mechanical eigenvalue problem
\begin{equation}
\left(-(\log\,q)^2\partial^2_\phi + \frac{q}{2}e^\phi\right) \Psi = \sqrt{1-q}(E-E_0)\Psi\,.
\end{equation}
This is equivalent to the Liouville form of the Schwarzian action in equation 32 in \cite{Bagrets:2017pwq}, given by
\begin{equation}
H=-{\partial^2_\phi \over 2M} + \gamma e^\phi
\label{eq:balgretsH}
\end{equation}
after a constant shift of $\phi$. In \cite{Bagrets:2017pwq}
$M$ was the scale $M=\frac{N\log N}{64J\sqrt{\pi}}$ (for the SYK model with quartic interactions). For us it is set by $ |\log(q)|^{-2} \sim \lambda^{-2}$.

The prescription in \cite{Bagrets:2016cdf,Bagrets:2017pwq} (and in \cite{Mertens:2017mtv} for 2D case) requires that, in the path integral, we sum over trajectories that begin and end in the strong coupling region $\phi\to\infty$. This is in qualitative agreement with our prescription since we place the state $v_0$ as initial and final states. Recall that $v_0=(1,0,0,0....)$, i.e., only the $i=0$ term is turned on, which where the term $q^i$ is the largest. In terms of $\phi$, $e^{\phi}$ is largest which is indeed the analogue of the Liouville strong coupling region. The models are of course not exactly the same since the model in \cite{Bagrets:2016cdf} captures the low energy and the ${\hat T}$ matrix captures the full dynamics.

This also gives an interpretation of the index $i$ via its relation to $\phi$. $\phi(t)$ measures where the AdS$_2$ space is glued to whatever non-universal UV we have (the leading effect being the Schwarzian action), i.e., $\phi(t)$ parametrizes the length of the AdS$_2$ throat. We see that in the full model the size of $AdS_2$ is actually quantized, giving rise to a minimal size $AdS$ which corresponds to the state $v_0$.

It is worth reiterating that the density of states for the $H$ Hamiltonian is different than the density of states for the matrix $\hat T$, even in the large $n$ limit. Rather the density of states in the former is related to the density of states in the latter $H$ by equation \eqref{DensToDens} or, equivalently, by
\begin{equation}
E\biggl(Tr_H(e^{-\beta H})\biggr) = v_0^\dagger e^{-\beta T} v_0 
\end{equation}
which means that we have to put a specific initial and final states for $\hat T$ in order to compute the partition function. It is tempting to interpret this in Minkowski space as a computation with initial and final states at the past and future singularities of the black hole.


\section{The two point function}
\label{sec:2 point function}

\subsection{The exact 2-pt function}

As explained in section \ref{sec:Motivation and Summary}, we want to compute correlators of random operators $M$ taken from the same universality class as the Hamiltonian \eqref{Hamilt}. Hence, these are defined by
\begin{equation}\label{RandOp}
M=3^{-p_m/2}{n\choose p_m}^{-1/2}\sum_J m_J \sigma_J\,,
\end{equation} 
where $J$ is now a string of $p_m$ distinct sites and Pauli matrices. The sum runs over all such possible $J$'s, and $m_J$ are independent Gaussian variables with zero mean and unit standard deviation (in particular they are also independent of the coefficients $\alpha_J$ in H). 

There are two relevant parameters that we will keep fixed in the limit $n\to \infty$.  The first is the analogue of $\lambda$ (see \eqref{ParamDef}) for the random operator \eqref{RandOp} 
\begin{equation}
	\lambda_m=\frac{3}{4}\frac{p_m^2}{n},\quad \quad q_m=e^{-\lambda_m}\,.
\end{equation}
The second is 
\begin{equation}
	{\tilde \lambda} = \sqrt{\lambda \lambda_m} = \frac{3}{4} \frac{pp_m}{n},\quad\quad {\tilde q}=e^{-\tilde\lambda}\,.
\end{equation}

We want to evaluate the exact thermal 2-pt function for the random operator M
\begin{equation}
	2^{-n} \cdot E\left[ \Tr(e^{-\beta H} M(t)M(0) ) \right], \ \ \text{or}\ \  2^{-n} \cdot E\left[ Tr(e^{-{\beta H \over 2}} M(t) e^{-{\beta H \over 2}} M(0) )  \right]
\end{equation}
for any value of $\beta$ and $t$. The formalism developed below, based on the set-up in section \ref{sec:new derivation}, proceeds 
by evaluating, and then resumming, expressions of the form
\begin{equation}\label{Trace2M}
2^{-n}\cdot  E\left[Tr(H^{k_1}MH^{k_2}M)\right]\, .
\end{equation}
This formalism can be extended to compute any n-pt function \cite{Vladi}.

The strategy is to reduce the computation to some relevant chord partition function, and then to evaluate it. The identification of the relevant partition function follows the discussion in section \ref{sec:new derivation}. The Gaussian integration over the random coefficients of the operators still pairs them. Hence, one can still think in terms of chord diagrams. The only difference is that the Gaussian integral over $m_J$'s pairs the two $M$ insertions, whereas the Gaussian integral over $\alpha_J$'s pairs the $k_1+k_2$ insertions of $H$. To take into account the index sets of the different chords and their intersections, one must evaluate the trace over the Pauli matrices. The arguments leading to only pairwise intersection and to the length of the intersections being Poisson distributed remain the same. The only difference is that intersections between two $H$-lines are determined by the length parameter of the $H$ operator, giving a factor of $q$ to their intersection, whereas the intersection of an $H$-line with $M$-line is determined by both the length of $M$ and the length of $H$, giving a factor of ${\tilde q}$ to their intersection. The net result is that one is left with a {\it marked chord diagram} where one chord is distinguished - an example is given in figure \ref{fig:A sample marked Chord diagram} - and the partition function that we are interested in is the marked chord partition function, as promised in section \ref{sec:Summary of results}.

To evaluate the marked chord diagram, we need to modify the "hopscotch" procedure described in section \ref{subsec:Evaluation of the Chord partition function} to include the marked chord. Given the distinguished nature of the pair of $M$ insertions, it is convenient to choose where to open the circular chord diagram in such a way that one $M$ appears to the rightmost of the line, and the other $M$ somewhere in the interior, as in figure \ref{fig:Chord diagram for Two point function}. The two operators $M$ are denoted by red dots and are paired by the bold faced line.  The procedure now consists in pairing the remaining $H$'s starting with the rightmost insertion of $H$. Propagating the system through the first $k_2$ steps, i.e the $k_2$ $H$'s between the two $M$ operators, gives the same contribution as before. However, when we hop over the second insertion of $M$, the open $H$-lines cross the $M$-line picking up an additional factor of ${\tilde q}^{\text{no.\ of\ lines}}$. The last step is to propagate for the remaining $k_1$ steps. The expression for the marked chord diagram, or the 2-pt function of M, is therefore

\begin{figure}[t!]
	\def\s{1.5}
	\begin{center}    
		\begin{tikzpicture}[scale=0.8,cap=round,>=latex]
		\foreach \x in {0,1,2,5,6,7,8,9}  
		{
			\filldraw[blue] (\x*\s,0) circle(2pt);
		}
		\foreach \x in {4,10} 
		{
			\filldraw[red] (\x*\s,0) circle(2pt);
		}
		\draw[ultra thick, color=black] (4*\s,0) -- (4*\s,2*\s) --  (10*\s,2*\s) --  (10*\s,0) ;
		
		\draw[thin] (0,1.2*\s) -- (2.2*\s, 1.2*\s) ;
		\draw[thin,dashed] (2.2*\s,1.2*\s) -- (5*\s, 1.2*\s) ;
		\draw[thin] (5*\s,1.2*\s) -- (6.8*\s, 1.2*\s) ;
		\draw[thin,color=black] (8*\s,0) arc (0:90:1.2*\s) ;
		
		\draw[thin] (0,1.1*\s) -- (2.2*\s, 1.1*\s) ;
		\draw[thin,dashed] (2.2*\s,1.1*\s) -- (4.9*\s, 1.1*\s) ;
		\draw[thin,color=black] (6*\s,0) arc (0:90:1.1*\s) ;	
		
		\draw[thin,color=black] (0*\s,0) arc (180:0:\s) ;

		\draw[thin,color=black] (1*\s,0) arc (180:90:0.9*\s) ;
		\draw[thin] (1.9*\s,0.9*\s) -- (2.2*\s, 0.9*\s) ;
		\draw[thin,dashed] (2.2*\s,0.9*\s) -- (5*\s, 0.9*\s) ;
		\draw[thin] (5*\s,0.9*\s) -- (8.1*\s, 0.9*\s) ;
		\draw[thin,color=black] (9*\s,0) arc (0:90:0.9*\s) ;

		\draw[thin,color=black] (5*\s,0) arc (180:90:0.8*\s) ;	
		\draw[thin] (5.81*\s,0.8*\s) -- (6.19*\s, 0.8*\s) ;
		\draw[thin,color=black] (7*\s,0) arc (0:90:0.8*\s) ;	
		
		\draw[ultra thick, color=black]
		{
			(3*\s,0*\s) node [right] { $ \dots $}
			(3*\s,13*\s/10) node [above] { $ \vdots $}
		};
		\end{tikzpicture}
		\caption{Marked chord diagram for a two point function. }
		\label{fig:Chord diagram for Two point function}
	\end{center}
\end{figure}
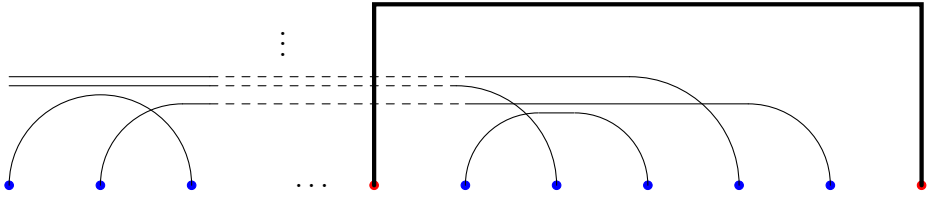
\begin{equation}
	2^{-n}\cdot E\left[ \Tr\bigl( H^{k_1}MH^{k_2}M\bigr) \right] = \langle 0 | T^{k_1} W({\tilde q}) T^{k_2} | 0 \rangle
\end{equation}
where 
\begin{equation}
	W({\tilde q}) = \text{Diag}(1,{\tilde q},{\tilde q}^2,....)
\end{equation}
encodes the intersection of an $H$-line with an $M$-line, when the former hops over the latter. 
Recalling that $T = P^{-1}\hat T P$ and noticing that $[P,W]=0$, we can also write the two point function as 
\begin{equation}\label{2PTHat}
	2^{-n}\cdot E\left[ \Tr\bigl( H^{k_1}MH^{k_2}M\bigr) \right] = \langle 0 | \hat T^{k_1} W({\tilde q}) \hat T^{k_2} | 0 \rangle
\end{equation}

The 2-pt function can be written in terms of the matrix elements of $\hat T$ defined in \eqref{eq:hat T matrix element} by inserting a complete set of states
\begin{equation}\label{2PTHat expanded}
\begin{split}
\langle 0 | \hat T^{k_1} W({\tilde q}) \hat T^{k_2} | 0 \rangle
& = \sum_{l=0}^\infty  \langle 0| {\hat T}^{k_1} | l \rangle   {\tilde q}^l  \langle l | {\hat T}^{k_2}| 0\rangle \\
& = \int d\theta_1 d\theta_2 \hat \psi_0(\theta_1)  \hat \psi_0(\theta_2) E(\theta_1)^{k_1} E(\theta_2)^{k_2} \sum_l \hat\psi_l(\theta_1) {\tilde q}^l {\hat\psi}_l(\theta_2) 
\end{split} 
\end{equation}
where $| l \rangle $ stands for a vector having 1 in the $l$'th place and $E(\theta) = {2 \cos \theta \over \sqrt{1-q}}$. Using the $\hat \psi$ wavefunctions in \eqref{eq:TEigV another expression}, the infinite sum \eqref{2PTHat expanded} reduces to 
\begin{eqnarray}\label{eq:Two Point function Matrix Element}
	\sum_l \hat\psi_l(\theta_1) {\tilde q}^l {\hat\psi}_l(\theta_2) &=& \hat\psi_0(\theta_1)  {\hat\psi}_0(\theta_2)
	\sum_l H_l(\cos(\theta_1) |q) H_l(\cos(\theta_2) |q) \frac{{\tilde q}^l}{(q;q)_l}  \\
\nonumber
	&=& \hat\psi_0(\theta_1)  {\hat\psi}_0(\theta_2)
	\frac{({\tilde q}^2;q)_\infty}{({\tilde q}e^{i(\theta_1+\theta_2)},{\tilde q}e^{i(-\theta_1+\theta_2)},{\tilde q}e^{i(\theta_1-\theta_2)},{\tilde q}e^{i(-\theta_1-\theta_2)};q)_\infty}
\end{eqnarray}
where we used the identity \eqref{eq:n ortthogonality with t}.

To evaluate $\mbox{Tr}[e^{-{\beta H\over 2}} M(t) e^{-{\beta H\over 2}} M(0)]$, we expand the exponentials insert \eqref{2PTHat}, and resum the power of the eigenvalues ${\hat T}$ matrices (which now appear twice) into an exponential.  One then gets
\begin{multline}\label{2PTTheta}
	E[\mbox{Tr}\left(e^{-{\beta H\over 2}} M(t) e^{-{\beta H\over 2}} M(0)\right)]  \\
	={(q;q)_\infty^2 ({\tilde q}^2;q)_\infty\over (2\pi)^2 }\int_0^\pi d\theta_1 d\theta_2 e^{{2cos(\theta_1)(-{\beta\over 2}+it)\over\sqrt{(1-q})}} e^{{2cos(\theta_2)(-{\beta\over 2}-it)\over\sqrt{(1-q})}}  \\
	\ \times { (e^{2i\theta_1},q)_\infty(e^{-2i\theta_1},q)_\infty(e^{2i\theta_2},q)_\infty  (e^{-2i\theta_2},q)_\infty  
		\over ({\tilde q}e^{i(\theta_1+\theta_2)},{\tilde q}e^{i(-\theta_1+\theta_2)},{\tilde q}e^{i(\theta_1-\theta_2)},{\tilde q}e^{i(-\theta_1-\theta_2)};q)_\infty}
\end{multline}
where the value \eqref{eq:TEigV another expression} of ${\hat\psi}_0(\theta)^2$ was inserted.

In the next subsection we will evaluate this expression for a special case of $q$ and $\tilde q$. But before we do that, we will perform a quick check on our results above. 

\paragraph{A check.} Before evaluating \eqref{2PTTheta} for a special case of $q$ and $\tilde q$, one can perform a check by taking the
$\tilde q \to 1$ limit. There should be no cost for the $H$ lines crossing the $M$ lines in this limit. Hence, it must be that
\begin{equation}
	E \left[ Tr(H^{k_1} M H^{k_2} M) \right] = E \left[ Tr(H^{k_1+k_2} ) \right] \quad \text{when} \quad \tilde q \to 1
\label{eq:check}
\end{equation} 
To check \eqref{2PTTheta} is compatible with this behaviour, notice that near $\tilde q \to 1$, $(\tilde q^2;q)_\infty \to 0$, due to the first term in the product. Hence \eqref{2PTTheta} vanishes, unless $\theta_1 \to \theta_2$, since an additional zero in the denominator occurs then\footnote{Another zero may appear in the denominator when $\theta_1+\theta_2=\pi$ but this appears in a co-dimension 2 in the range of integration.}. Hence, the integrand in \eqref{2PTTheta} behaves like a delta function whose strength is given by 
\begin{equation}
	{ (\tilde q^2 ; q)_\infty \over |(\tilde q e^{i(\theta_1-\theta_2)};q)_\infty|^2    } = {2 (1-\tilde q)   \over (1-\tilde q)^2 + (\theta_1-\theta_2)^2} \times { (q;q)_\infty\over (q;q)_\infty^2} = 2\pi \delta(\theta_1-\theta_2) \times {1 \over (q;q)_\infty}
\end{equation}
Thus, in the $\tilde{q}\to 1$ limit, the correlator \eqref{2PTTheta} equals 
\begin{equation}
	{ (q;q)_\infty   \over (2\pi) }
	\ \int d\theta_1 E^{k_1+k_2}(\theta_1)   \  |(e^{2i\theta_1};q)_\infty|^2  = \int d\theta E^{k_1+k_2}(\theta) \Psi(\theta,q) = m_{k_1 + k_2}(q)\,,
\end{equation}
where we used \eqref{eq:Our moment formuale} in the last step, in agreement with \eqref{eq:check}.

\subsection{The $q \to 1$ limit with ${\tilde q}=q^m$}

The exact 2-pt function of $M$ \eqref{2PTTheta} holds for all ranges of time (which are held fixed in the $n\to\infty$ limit). In the remainder of this section we will compute the formula in a specific case, which is the low energy regime where conformal symmetry is expected to appear, as discussed in section \ref{sec:qeq1 limit} and in Appendix B of \cite{Cotler:2016fpe}.

We will work in the limit $q,\ {\tilde q}\to 1$ since we want to work in the limit in which the correlators of each operator separately approximately factorize. However, more and more terms contribute in the Pochammer symbols in this limit, similar to what we had for the partition function, and hence it is important how we take this limit. Since in gravity non-factorization of correlation functions for different operators is governed by the same parameters (e.g. the same 1/N), then the rates of $q\to 1$ and ${\tilde q}\to 1$ should be related. 

A particularly simple case to analyze is ${\tilde q}=q^m$ with $m$ an integer. This has technical advantages, but it is also physically interesting because it corresponds to 
\begin{equation}
	p_m=m\ p\,.
\end{equation}
That is, if the Hamiltonian is made out of a sum of strings of p spin operators (with random coefficients), then the random operator M is made out of a string of $m\cdot p$ spin operators. This is reminiscent of the statement that, say for $4D,\ {\cal N}=4$ SYM, the Hamiltonian is a descendant of $\text{Tr}(X^2)$, yet we can probe the system with low energy fields, which correspond to single trace operators of the form $\text{Tr}(X^n)$, $n>2$, and their conformal descendants. 

As discussed in \cite{Cotler:2016fpe} and section \ref{sec:qeq1 limit}, our model has a conformal low energy limit. Hence, conformal symmetry should assign a dimension one to the Hamiltonian. If the fundamental fields (in this case the spin operators) can be assigned a specific conformal dimension, and if this conformal dimension is additive in composite operators - as in the SYK model on both counts - then one expects the conformal dimension of $M$ to be $m$. We will see how our exact formula matches this, up to the existence of mixing with operators of lower dimension when we work at finite temperature. Despite this, our exact formula always has an overlap with an operator of the right dimension.

Before doing the computation we would like to recall an additional formula to which we will compare our result. We will actually be computing the ``two sided correlator" 
\begin{equation}
	2^{-n} E\left[ Tr\bigl( e^{-{\beta H \over 2}} M(t) e^{-{\beta H \over 2}} M(0) \bigr) \right]. 
\end{equation}
This computation is slightly easier than the ordinary thermal correlator. We refer to this as the two sided correlator since, in an eternal black hole in $AdS$, it is the relevant correlator when there is one operator on each of the boundaries. For an particle of mass $M$ in the BTZ black hole this correlator is (see for example  \cite{Shenker:2013pqa} in the Eikonal approximation, with a shock wave there)
\begin{equation}
	\propto {1\over \cosh\bigl(\pi\beta (t_R-t_L)\bigr)^{2Ml}}
\end{equation}
where $l$ is the $AdS_3$ radius, $M$ is the mass of the particle and $Ml$ is the conformal dimension of the associated operator. This is what is expected from conformal invariance. In the single sided correlator the cosh is replaced by a sinh, to obtain the expected short distance behaviour of $1/t^{2Ml}$, and the correct Euclidean time periodicity.

\subsubsection{The reduced formula}

When $\tilde q  =q^m$, the identity 
\begin{equation}
	|(\tilde q e^{i \phi} ; q )_\infty |^2 = { |( e^{i \phi} ; q )_\infty|^2 \over  4 \sin^2 {\phi \over 2}\prod_{l=1}^{m-1} |1-e^{i \phi} q^l|^2 }\,,
\end{equation} 
allows to write the 2-pt function \eqref{2PTTheta} as
\begin{equation}\label{2PTB}
	\begin{split}
		16 (q^{2m};q)_\infty & {(q;q)_\infty^2 \over (2\pi)^2}  \times \int _0^\pi d\theta_1 d\theta_2 e^{{2cos(\theta_1)(-{\beta\over 2}+it)\over\sqrt{(1-q})}} e^{{2cos(\theta_2)(-{\beta\over 2}-it)\over\sqrt{(1-q})}} 
		\sin^2({\theta_1 + \theta_2 \over 2}) \sin^2({\theta_1 - \theta_2 \over 2})\\
		& \times \prod_{l=1}^{m-1} |(1-q^{l} e^{i(\theta_1+\theta_2)})|^2 |(1-q^{l} e^{i(\theta_1-\theta_2)})|^2
		\times 
		{ |(e^{2i\theta_1},q)_\infty|^2 |(e^{2i\theta_2},q)_\infty^2 
			\over   |(e^{i(\theta_1+\theta_2)};q)_\infty|^2 |(e^{i(\theta_1-\theta_2)};q)_\infty|^2  }
	\end{split}
\end{equation}
Notice the finite product can be rewritten, within the integral, as 
\begin{equation}\label{eq:Define D operator}
	\begin{split}
		{\cal D}_m(t,\beta) \equiv & \prod_{l=1}^{m-1} |(1-q^{l} e^{i(\theta_1+\theta_2)})|^2 |(1-q^{l} e^{i(\theta_1-\theta_2)})|^2 \\
		= & \prod_{l=1}^{m-1}  \bigg[ (1-q^{2l})^2 + q^l \left({1+q^l \over 2}\right)^2 (1-q) (-i\partial_t)^2 -  q^l (1-q^l)^2 (1-q) (\partial_\beta)^2 \bigg]  \\
		 \equiv & \prod_{l=1}^{m-1} D_l(t,\beta) \xrightarrow{q\to 1} (1-q)^{m-1} (-i\partial_t)^{2m-2}\ .
	\end{split}
\end{equation}
Taking the derivatives outside of the integral, allows to write the integrand in terms of $\vartheta$ functions (see \eqref{Theta-poch}) depending only on $q$, 
\begin{equation}
	\begin{split}
		16 (q^{2m};q)_\infty & {(q;q)_\infty^2 \over (2\pi)^2}   \ \  {\cal D}_m(t,\beta) \int _0^\pi d\theta_1 d\theta_2 e^{{2cos(\theta_1)(-{\beta\over 2}+it)\over\sqrt{(1-q})}} e^{{2cos(\theta_2)(-{\beta\over 2}-it)\over\sqrt{(1-q})}} 
		\\
		& \times \sin({\theta_1 + \theta_2 \over 2}) \sin({\theta_1 - \theta_2 \over 2})    \sin \theta_1 \sin \theta_2 \ \  {\vartheta_1({\theta_1 \over \pi } | {i \lambda \over 2\pi}) \vartheta_1({\theta_2 \over \pi } | {i \lambda \over 2\pi})  \over  \vartheta_1({\theta_1 + \theta_2 \over 2\pi } | {i \lambda \over 2\pi}) \vartheta_1({\theta_1-\theta_2 \over 2\pi } | {i \lambda \over 2\pi}) }
	\end{split}
\end{equation}
The $q \to 1$ limit of the Jacobi Theta functions is evaluated as in \eqref{eq:qeq1 of Jacobi Theta}, bringing the 2-pt function to the form
\begin{equation} 
	\begin{split}
		32 (q^{2m};q)_\infty & {(q;q)_\infty^2 \over (2\pi)^2}   \ \  {\cal D}_m(t,\beta)  \int _0^\pi d\theta_1 \int_0^\pi d\theta_2 e^{{2cos(\theta_1)(-{\beta\over 2}+it)\over\sqrt{\lambda}}} e^{{2cos(\theta_2)(-{\beta\over 2}-it)\over\sqrt{\lambda} }}  
		\\
		& \times \sin({\theta_1 + \theta_2 \over 2}) \sin({\theta_1 - \theta_2 \over 2})    \sin \theta_1 \sin \theta_2\ \  e^{-{3\pi^2 \over {2} \lambda}  -{1\over \lambda} {(  \theta_1  -{ \pi \over 2} )^2} -  {1\over \lambda}  {( \theta_2 -{ \pi \over 2} )^2}  }\\
		& \times  { \sinh({2 \pi \theta_1\over \lambda}) \sinh({2\pi(\pi-\theta_1)\over  \lambda})  \over \sinh({ 2\pi ({\theta_1+\theta_2\over 2})\over \lambda}) \sinh({ 2\pi (\pi - {\theta_1+\theta_2\over 2})\over \lambda})  } \ { \sinh({2 \pi \theta_2\over \lambda})  \sinh({2 \pi (\pi-\theta_2) \over \lambda})  \over \sinh({ 2\pi ({\theta_1-\theta_2\over 2})\over \lambda}) (1-2\cosh({2\pi(\theta_1-\theta_2)\over \lambda })e^{-4\pi^2/\lambda}  )  }  \\
		& \equiv
		{32} (q^{2m};q)_\infty  {(q;q)_\infty^2 \over (2\pi)^2}   \ \  {\cal D}_m(t,\beta) \cdot I(\beta,t,q)
	\end{split}
\label{eq:exact2pt}
\end{equation}
This is the exact 2-pt function for $q=\tilde{q}^m$ in the limit $\lambda\to 0$. In the next subsections, we study the function $I(\beta,t,q)$, from which all $m>1$ correlators can be extracted, in the low temperature and very low temperature regimes (or long time, and very long time regimes).

\subsection{Low and very low temperature regimes}

Since the integral $I(\beta,t,q)$ localizes near the edges at low energies, we define $\phi_i=\pi-\theta_i$. Expanding the integral near $\phi_i\sim 0$, 
\begin{equation} 
	\begin{split}
		I(\beta,t,q)= 
		&\ {1\over 8} e^{2\beta\over\sqrt{\lambda}} \int _0^\pi d\phi_1 \int_0^\pi d\phi_2 \,e^{{\phi_1^2(-{\beta\over 2}+it-{1\over \sqrt{\lambda}})\over\sqrt{\lambda}}} e^{{\phi_2^2(-{\beta\over 2}-it-{1\over\sqrt{\lambda}})\over\sqrt{\lambda} }}  
		\\
		& \times (\phi_1+\phi_2)(\phi_2-\phi_1)\phi_1\phi_2  
		\times  { \sinh({2\pi \phi_1\over  \lambda})  \over  \sinh({ 2\pi ( {\phi_1+\phi_2\over 2})\over \lambda})  } { \sinh({2 \pi \phi_2 \over \lambda})  \over \sinh({ 2\pi ({\phi_2-\phi_1\over 2})\over \lambda})  }  \\
	\end{split}
\label{eq:step1}
\end{equation}
As explained in section \ref{sec:thermo}, the low energy (and very low energy) regime satisfies $\beta\,\sqrt{\lambda} \gg 1$. To study the behaviour of the Gaussian factors in the above integral, it is convenient to rescale the integration variable $\varphi_i\equiv {\phi_i \over \lambda}$ together with the time and temperature parameters ${\tilde\beta}\equiv\lambda^{3/2}\beta,\ {\tilde t}\equiv \lambda^{3/2} t$. The low energy regime is equivalently described by
\begin{equation}
	\beta\,\sqrt{\lambda} \gg 1 \quad \Leftrightarrow \quad  {\tilde\beta} \gg \lambda
\end{equation}
allowing to approximate \eqref{eq:step1} by
\begin{equation}\label{eq:Two Point Function I}
	\begin{split}
		I(\beta,t,q) = & {\lambda^6\over 8} e^{2\beta\over\sqrt{\lambda}} \int _0^{\pi/\lambda} d\varphi_1 \int_0^{\pi/\lambda} d\varphi_2 \,e^{{\varphi_1^2(-{{\tilde\beta}\over 2}+i{\tilde t})}} e^{{\varphi_2^2(-{{\tilde\beta}\over 2}-i{\tilde t}) }}  
		\\
		& \times (\varphi_1+\varphi_2)(\varphi_2-\varphi_1)\varphi_1\varphi_2  
		\times  { \sinh({2\pi \varphi_1})  \over  \sinh({ 2\pi ( {\varphi_1+\varphi_2\over 2})})  } { \sinh({2 \pi \varphi_2 })  \over \sinh({ 2\pi ({\varphi_2-\varphi_1\over 2})})  }  \\
	\end{split}
\end{equation}

This integral has two regimes, following a similar discussion for the partition function in section \ref{sec:thermo} : 
\begin{itemize}
	\item The low energy-long time regime characterised by ${\tilde\beta},\, {\tilde t} \ll 1$, where the integral receives contributions from the range $\varphi_i \gg 1$. 
	\item The very low temperature regime, or very long time scale regime, characterised by ${\tilde \beta} \gg 1$ or ${\tilde t}\gg 1$, where the integral receives contributions primarily from $\varphi_i \ll 1$.
\end{itemize}

\subsubsection{Low temperature regime}

The low energy-long time regime ${\tilde\beta}, {\tilde t} \ll 1$ allows to extend the range of integration to $\infty$ since the gaussian in the integrand cuts off the integral well before the limits in \eqref{eq:Two Point Function I}. Notice also the integral is supported at large values of $\varphi_1, \varphi_2$, allowing us to approximate three of the $\sinh$ functions by their larger exponentials
\begin{equation}
	\begin{split}
		I(\beta,t,q )& =  {\lambda^6 e^{2\beta\over\sqrt{\lambda}} \over 2}  \int_0^\infty d\varphi \int_{-\varphi}^\varphi d\sigma  e^{-{\tilde\beta} \varphi^2 - {\tilde\beta}\sigma^2 - 4i{\tilde t} \varphi\sigma}  
		\varphi \sigma (\varphi^2 - \sigma^2) {e^{2\pi\varphi}\over \sinh{2\pi \sigma}}
\end{split}
\end{equation}
where we changed variables to $\varphi={\varphi_1+\varphi_2 \over 2},\ \sigma={\varphi_2-\varphi_1 \over 2}$. Due to the ${1 \over \sinh(2\pi\sigma)}$ term, the $\sigma$ integral receives contributions from finite $\sigma$, whereas its limit of integration is $\pm\varphi$, much larger quantities. This means we can trade the $\sigma$ limits by $\pm\infty$. Furthermore, we can also neglect the $e^{-{\tilde\beta}\sigma^2}$ term and the $\sigma^2$ term relative to $\varphi^2$ in the $(\varphi^2-\sigma^2)$ term. After these approximations, our integral reduces to
\begin{equation}
	\begin{split}
		I = {\lambda^6 e^{2\beta\over \sqrt \lambda} \over 2}  \int_0^\infty d\varphi  e^{-{\tilde\beta} \varphi^2 } \varphi^3 e^{2\pi\varphi}  \int_{-\infty}^\infty d\sigma e^{ - 4i{\tilde t} \varphi\sigma}  
		{\sigma\over \sinh{2\pi \sigma}}
	\end{split}
\end{equation}
Using the identity
\begin{equation}
		\int_{-\infty}^{\infty} d\sigma e^{-{4 i {\tilde t}\sigma \varphi } }    \frac{\sigma}{\sinh(2 \pi \sigma)}   =   \frac{1}{8 \cosh^2( {\tilde t} \varphi)},
\end{equation}
and introducing a further variable of integration $\varphi \equiv {\varphi_s\over \sqrt{ \tilde \beta}}+{\pi \over \tilde \beta}$, we finally get
\begin{equation}
\begin{aligned}
  I(\beta,t,q)&= {\lambda^6 e^{2\beta\over \sqrt \lambda} \over 16} \int_0^\infty d\varphi { \varphi^3  e^{-{\tilde\beta}\varphi^2+2\pi \varphi}\over  \cosh^2{({\tilde t}\varphi)}} \\ 
  &= {\pi^3 \lambda^6 e^{{2 \beta \over \sqrt \lambda}+ {\pi^2 \over \tilde \beta} }\over 16 \tilde \beta^3 \sqrt{\tilde \beta}}  \int_{- {\pi \over \sqrt{\tilde \beta}}}^\infty d\varphi_s e^{-\varphi_s^2}{ (1 + {\sqrt{\tilde \beta} \varphi_s \over \pi})^3 \over   \cosh^2{  [{{\pi\tilde t}\over \tilde \beta} (1+ {\sqrt{\tilde \beta} \varphi_s \over \pi} )]} }
\end{aligned}
\label{eq:phi saddle}
\end{equation}
Since ${\tilde\beta}\ll 1$ we can in any case neglect the $\varphi_s$ dependence in the numerator. The integral shows different behaviours depending on the scaling of ${\tilde t  }$ :

\begin{itemize}
	
	\item  When ${ \tilde t \over \sqrt{\tilde \beta} } \ll 1 $, the $\varphi_s$ dependence in the denominator can be neglected and, to leading order in $\tilde\beta$, the result is
\begin{equation}\label{LowTRegA} 
	I(\beta,t,q) = { \lambda^{3 \over 4} \over 16}  \left({\pi \over \beta}\right)^{7 \over 2} e^{{2 \beta \over \sqrt \lambda}+ {\pi^2 \over \beta \lambda^{3 \over 2}} }  \ {1 \over \cosh^2\left( \pi t \over \beta\right)}\,, \quad \quad 1\gg \sqrt{\tilde{\beta}}\gg \tilde{t}
\end{equation}
	
	\item When $ \tilde t \gg \sqrt{\tilde\beta}(\gg{\tilde\beta})$, the cosh in the denominator contributes. Keeping only the larger exponential due to the latter,  the integral becomes
\begin{equation}\label{LowTRegB}
\begin{aligned}
	I(\beta,t,q) &=		{\lambda^6 e^{2\beta\over \sqrt{\lambda} } \over 4} \int_0^\infty d\varphi e^{-{\tilde\beta}\varphi^2+2\pi \varphi-2{\tilde t} \varphi} \varphi^3\\
	& = {\lambda^6 e^{2\beta\over \sqrt{\lambda} } \over 4}  e^{{(\pi-{\tilde t})^2\over {\tilde\beta}}} {(\pi -{\tilde t})^3\over {\tilde\beta}^3 \sqrt{\tilde \beta}}\int_{-{\pi - \tilde t \over \sqrt{\tilde \beta}}}^\infty d\varphi_s \,e^{-\varphi_s^2} \bigg( 1+ {\sqrt{\tilde \beta} \varphi_s \over \pi - \tilde t } \bigg)^3 \\
	& \sim \frac{\lambda^{3/4}}{16}\left(\frac{\pi}{\beta}\right)^{7\over 2}  e^{2\beta\over \sqrt{\lambda}}\,4e^{\frac{(\pi-{\tilde t})^2}{\tilde\beta}}\,, \quad \quad 1\gg \tilde{t} \gg \sqrt{\tilde\beta}
\end{aligned} 
\end{equation}
where we changed the integration variable to $\varphi \equiv { \varphi_s \over \sqrt{ \tilde\beta} } + { \pi-{\tilde t} \over  \tilde\beta }$ in the second step. Due to the large $t/\beta$, or ${\tilde t}/{\tilde \beta}$, \eqref{LowTRegB} differs from \eqref{LowTRegA} by an additional $e^{{\tilde t}^2\over {\tilde\beta}}$.
		
\end{itemize}

\subsubsection{Very low temperature}

When $\tilde \beta = \beta\lambda^{3/2} \gg 1$, the angles $\varphi_1$ and $\varphi_2$ are localized to a range much smaller than $1$. This allows to expand the $\sinh$ functions in \eqref{eq:Two Point Function I} to obtain

\begin{equation} \label{eq:Two Point function partial very low temp}
	I(\beta,t,q) =  {\lambda^6 \over 2} e^{2 \beta \over \sqrt \lambda} \int _0^\infty d\varphi_1 \int_0^\infty d\varphi_2 \,e^{ \varphi_1^2(-{\tilde \beta \over 2}+i{\tilde t}) +\varphi_2^2(-{\tilde \beta \over 2}-i{\tilde t}) }  \varphi_1^2\varphi_2^2  =  {\pi \lambda^6 e^{2 \beta \over \sqrt \lambda} \over 4(\tilde \beta^2 + 4 \tilde t^2)^{3 \over 2}}
\end{equation}
where we traded the upper limit with $\infty$. This may have the following interpretation. This quantity equals
\begin{equation}
	\int dE_1 dE_2 \rho(E_1)\rho(E_2) e^{-\beta_1 E_1-\beta_2 E_2} E\bigg(|\langle E_2 |M |E_1\rangle |^2\bigg)
\end{equation}
where $E( )$ is the statistical average and $\beta_1,\beta_2$ are related to $\beta,t$.  This means that
\begin{equation} \label{LowERnd}
	E\bigg(|\langle E_2 |M |E_1\rangle |^2\bigg)\sim \varphi_1^2\varphi_2^2 \sim E_1 E_2
\end{equation}
where $E_i$ measures the energy of the state above the ground state. We can interpret this as if the operator $M$ acts as an underlying gaussian random matrix which couples to low energy states with form factors $\varphi^2$, i.e. consider a set of random vectors 
\begin{equation}
	|v_\alpha \rangle = \sum_{i} \sqrt{E_i} c_{i,\alpha} |E_i\rangle, 
\end{equation}
where the sum is up to some energy higher than the scale set by the very low temperature, and $c_{i,\alpha}$ are independent complex Gaussian random variables with mean zero and unit standard deviation. Take $M$ to be a random Gaussian Hermitian matrix in terms of these variables
\begin{equation}
	M=\sum_{\alpha,\beta} |v_\alpha\rangle {\hat M}_{\alpha,\beta} \langle v_\beta |
\end{equation}
where  and ${\hat M}$ are independent complex Gaussian variables with mean zero and standard deviation 1. In this case \eqref{LowERnd} is satisfied.

\subsection{ The final correlator}

The evaluation of the exact 2-pt function \eqref{eq:exact2pt} requires to compute the action of the operator ${\cal D}_m(t,\beta)$ in \eqref{eq:Define D operator} on $I(\beta,t,q)$ and interpret the result. It is easy to read the results without actually having to worry about the details of ${\cal D}_m(t,\beta)$.

\begin{itemize}
	
	\item The low temperature regime corresponds to the conformal regime, when the fluctuations of the pseudo-Goldstone modes are still small. Assigning the Hamiltonian $H$ the conformal dimension 1, one would expect an operator made of $m$ spin operators to have dimension $m$.   
	
	This is exactly what happens in our formulas. For $m=1$, the operator ${\cal D}_1(t,\beta)$ reduces to the identity. Hence, our result \eqref{LowTRegA} is the correlator for an operator of dimension 1 i.e., $\sim { 1 \over \cosh^2({\pi t \over \beta})}$.
	
	For $m>1$, there exists operator mixing, but we can extract the operator content from the correlator as follows. To isolate the conformal dimensions of the participating operators, first insert the operators on the same side, or equivalently take $t=i\beta/2+t'$. This turns the cosh into a sinh. Second, take the limit $t' \ll \beta$. In this case the leading contribution\footnote{This can be seen from \eqref{eq:Define D operator}. More precisely $\partial_t$ appears as $\lambda\partial_t^2$ and $\partial_\beta$ appears as $\lambda^3\partial_\beta^2$. Acting with them on a function of the form ${\beta^A \over t^B}$ we obtain an expression $({\lambda^3 \over \beta^2})^{n_\beta}({\lambda \over t^2})^{n_t} {\beta^A \over t^B}$. For a fixed power of $t$, an addition derivative with respect to $\beta$ adds a power of ${\lambda^3 \over \beta^2} \ll 1$. }  
	in ${\cal D}_m(t,\beta)$ acts on it with $\partial_t^{2m-2}$ turning the correlator into $1/t^{2m}$ which is the 2-pt function for an operator of dimension m.
	
	\item For the very low temperature/long time regime we can compare \eqref{eq:Two Point function partial very low temp} with equation (67) in \cite{Bagrets:2017pwq}. Although their discussion is for SYK model with quartic interactions, it is within the Liouville description of the Schwarzian action. Since our spin glass model reproduces the latter in this very low temperature regime, both results should be similar. The finite temperature 2-pt function of a pair of Majorana fermions in the SYK model at long times/low temperatures (in the conventions used in \cite{Bagrets:2017pwq}) equals 
	\begin{equation}
		G(\tau) \sim - \frac{M^2\beta^{1/2}}{\sqrt{J}}\,\frac{\text{sgn}(\tau)}{\tau^{3/2}(\beta - \tau)^{3/2}}\,, \quad \quad \tau\gg M \equiv \frac{N\log N}{64\sqrt{\pi}\,J}
		\label{eq:altland}
	\end{equation}
	For a 2-pt function of higher dimension operators, the time dependence (at long time and low temperature) remains with the same power, except that the coefficient of power of $M$ in front of the expression increases.   
	
	To match with \eqref{eq:Two Point function partial very low temp}, one needs to work with Lorentzian time, $\tau=it$ and to shift the time imaginary axis by $t\to t -i\frac{\beta}{2}$. Altogether, 
	\begin{equation*}
		\tau^{3/2}(\beta - \tau)^{3/2} \to \left(\frac{\beta^2}{4}-\tau^2\right)^{3/2} \to \left(\frac{\beta^2}{4}+ t^2\right)^{3/2}\,.
	\end{equation*}
	where we analytically continued back to lorentzian time in the last step. 
	
	This is to be compared with our expression following from \eqref{eq:Two Point function partial very low temp}
	\begin{equation}
		G^{(m)}(t) = - { 2 \over \pi} (q^{2m};q)_\infty (q;q)^2_\infty {\cal D}_m(t,\beta) \frac{\lambda^{3 \over 2}\,e^{2\beta \over \sqrt{\lambda}}}{4 \left(\frac{\beta^2}{4}+ t^2\right)^{3/2}}\,.
	\end{equation}
	in the limit ${\beta^2 \over 4}+t^2\to\infty$. For $m=1$, the $t$ dependence agrees. For $m>1$, the leading long time behaviour also agrees, and arises from the terms in ${\cal D}_m(t,\beta)$ which either have no derivatives, or have the $\partial_\beta^2$ terms acting on the $e^{2\beta/\sqrt{\lambda}}$. Note, however, that these are not strictly reliable results as this expression is multiplied by $\lambda^{2(m-1)}$, and we dropped terms of similar order throughout our discussion. However, within the terms that we kept, the leading very low temperature/very long time expressions agrees with that of \cite{Bagrets:2017pwq}.

\end{itemize}

\vspace{20pt}
\subsection*{Acknowledgements}
We would like to thank O. Aharony, A. Altland, D. Bagrets, M. Isachenkov, N. Itzhaki, D. Kutasov, V. Narovlansky, M. Rozali, S. Shenker and G. Torrents for useful discussions. The work of MB is supported by an ISF center of excellence grant (1989/14). The work of JS is supported by the Science and Technology Facilities Council (STFC) [grant number ST/L000458/1]. PN is grateful for support provided by International Centre  for Theorectical Sciences, India where part of this work was done. MB holds the Charles and David Wolfson Professorial chair of Theoretical
Physics.


\appendix{}

\section{q-Pochammer symbols and Jacobi $\vartheta$ functions}
\label{sec:Theta functions}

The q-Pochammer symbol is defined as  
\begin{equation}
  (a;q)_n \equiv \prod_{k=0}^{n-1} (1-a\,q^k)\,.
\label{q-poch}
\end{equation}
It allows an extension to an infinite product
\begin{equation}
  (a;q)_\infty \equiv \prod_{k=0}^{\infty} (1-a\,q^k)\,.
\label{qinf-poch}
\end{equation}
The Jacobi $\vartheta_1$ function is defined as (see equation (8.A.2) in \cite{GSW-2})
\begin{equation}\label{def-theta1}
\vartheta_1(\theta | \tau) \equiv  2 \bar q^{1\over 4} \sin \pi\theta \prod_{m=1}^\infty [ (1-\bar q^{2m}) (1 - 2 \cos(2 \pi \theta)  \bar q^{2m} + \bar q^{4m}) ]\,, \quad \text{with} \quad \bar q =e^{i \pi \tau}
\end{equation}
It is convenient for our discussion in section \ref{sec:qeq1 limit} to write products of q-Pochammer symbols in terms of the Jacobi $\vartheta_1$ function. To do this, note that with $q = e^{-\lambda}$, 
\begin{equation}\label{Theta-poch}
\begin{split}
\vartheta_1({\theta \over \pi}| {i \lambda \over 2\pi})
& = 2 q^{1 \over 8} \ \sin\theta \ (q; q)_\infty (e^{2 i \theta } q; q)_\infty 
(e^{-2 i\theta } q; q)_\infty \\
& =  { q^{1 \over 8} \over 2 \sin\theta} (q; q)_\infty (e^{2 i \theta } ; q)_\infty 
(e^{-2 i\theta } ; q)_\infty
\end{split} 
\end{equation}
\paragraph{Modular Transformations of Jacobi Theta functions.} $\vartheta$ functions obey modular transformation properties. The one relevant for our purposes is (see 8.A.20 in \cite{GSW-2})
\begin{equation}\label{eq:Modular Transformation Formulae}
  \vartheta_1(z |\tau) = \vartheta_1( -{z \over \tau} | -{1 \over \tau} ) \frac{1}{\eta\,(\tau)^{1\over 2}} e^{-i \pi z^2/\tau}\,, \quad \text{with} \quad \eta=e^{i\pi/4}\,.
\end{equation}

\section{Some properties of Continuous q-hermite polynomials}
\label{section Appendix:Some properties of Continuous q-hermite polynomials}

Continuous q-hermite polynomials are defined as
\begin{equation}
H_n(x|q) \equiv \sum_{k=0}^n  { (q;q)_n \over (q;q)_k (q;q)_{n-k}  } e^{i(n-k) \theta}\,, \hspace{20mm} x = \cos \theta
\end{equation}
They can equivalently be defined using the generating function\footnote{ $(a_1,a_2,\dots;q)_\infty \equiv (a_1;q)_\infty (a_2;q)_\infty \dots $}.
\begin{equation}
\sum_{n=0}^\infty H_n(x|q)  {t^n \over (q;q)_n } = {1 \over (t e^{i \theta}, t e^{-i \theta} ;q)_\infty }
\end{equation}
$H_n(x|q)$ turns out to be a polynomial in $x,q$. We list its relevant properties for our work below (see \cite{gasper2004basic} for example):

\begin{itemize}
\item $H_n(x|q)$ satisfies a recursion relation (wikipedia)
\begin{equation}\label{eq:Recursion of H_n}
2 x H_n(x|q) = H_{n+1}(x|q) + (1-q^n) H_{n-1}(x|q) \hspace{10mm} H_{-1}(x|q) =0, H_{0}(x|q) =1,
\end{equation}

\item $H_n(x|q)$ satisfies 
x-orthogonality of the form
\begin{equation}\label{eq:x-orthogonality}
\int_0^\pi H_{m}(\cos \theta | q) H_{n}(\cos \theta | q) |(e^{2i\theta} ; q)_\infty|^2 d\theta= 2 \pi \ {(q;q)_n \over (q;q)_\infty } \delta_{mn}
\end{equation}

\item $H_n(x|q)$ satisfies 
n-orthogonality of the form
\begin{equation}\label{eq:n ortthogonality with t}
\sum_{n=0}^\infty H_n(x|q) H_n(y|q) {t^n \over (q;q)_n } = { (t^2 ; q)_\infty \over (t e^{i (\theta+\phi)}, t e^{i (\theta-\phi)},t e^{-i (\theta-\phi)},t e^{-i (\theta+\phi)} ;q)_\infty } 
\end{equation}
To normalize the eigenfunctions of $\hat T$ in section \ref{eq:eigensystem} (see \eqref{TEigV}), we need the $t \to 1$ limit of the above identity. In this limit, the right hand side above can be expanded as 
\begin{equation}
\begin{aligned}
  \frac{(t^2 ; q)_\infty}{|(t e^{i (\theta+\phi)};q)_\infty|^2 |(t e^{i (\theta-\phi)};q)_\infty|^2} &= \frac{(1-t^2)}{|1-t e^{i (\theta+\phi)}|^2 |1-t e^{i (\theta-\phi)}|^2} \\
  & \cdot \frac{(qt^2;q)_\infty}{|(qt e^{i (\theta+\phi)};q)_\infty|^2 |(qt e^{i (\theta-\phi)};q)_\infty|^2}
\end{aligned}
\end{equation}
Since this expression vanishes for $t=1$ at generic values of $\theta,\phi$, but blows up for $\theta=\pm \phi$, it is proportional to $\delta(\theta\pm\phi)$. To determine the strength of the $\delta$ function, one notes that if  $t\equiv 1-\epsilon$, in the $ \epsilon  \to0, \theta \to \phi$ limit the expression above becomes\footnote{We need to use $  \delta (t) = \lim_{\epsilon\to 0} \frac{1}{\pi} \frac{\epsilon}{\epsilon^2 + t^2}$ and the identity $(q\,e^{2i\theta};q)_\infty = \frac{(e^{2i\theta};q)_\infty}{1-e^{2i\theta}}$}
\begin{equation}
  \frac{2\epsilon}{(\epsilon^2+ (\theta-\phi)^2)}  \frac{(q,q)_\infty}{|(  e^{2i \theta};q)_\infty|^2 |(q ;q)_\infty|^2}
 = 2 \pi \delta(\theta-\phi) \frac{1}{|(  e^{2i \theta};q)_\infty|^2 |(q ;q)_\infty|}\,,
\end{equation}
%
Hence, the $t=1$ limit of the formulae \eqref{eq:n ortthogonality with t} reads
\begin{equation}
\sum_{n} {H_n(x|q) H_n(y|q)  \over (q;q)_n } =   \frac{2\pi \left(\delta(\theta-\phi)  + \delta(\theta+\phi)\right)}{|( e^{2i \theta};q)_\infty|^2 (q ;q)_\infty} 
\end{equation}
\end{itemize}

Using these identities, one can fix the normalization constant in \eqref{TEigV}. Define  
\begin{equation}\label{eq:hat T Eigenvector}
\hat \psi_l(x|q) \equiv \sqrt{(q ;q)_\infty} |(   e^{2i \theta};q)_\infty|  \ \ \frac{H_l( x |  q )}{\sqrt{2\pi (q;q)_l}} 
\end{equation}
The latter satisfies both the unit normalized n-orthogonality and $x$-orthogonality relation\footnote{The $n$-orthogonality relation also has a $\delta(\theta+\phi)$ term, which we will neglect below since they give negligible contribution in $\int d\theta  d\phi$ compared to $\delta(\theta-\phi)$.}
\begin{eqnarray}
\sum_{n=0}^\infty  \hat \psi_n(\cos\theta|q) \hat \psi_n(\cos \phi|q)  & = &  \delta(\theta-\phi)    \\ 
\int_0^\pi \hat \psi_{m}(\cos \theta | q) \hat \psi_{n}(\cos \theta | q) d\theta &=  & \delta_{mn}
\end{eqnarray}
Finally, note that \eqref{eq:hat T Eigenvector} can also be rewritten as 
\begin{equation}\label{eq:TEigV another expression}
 \psi_l(x|q) =  \psi_0(x|q)   \frac{H_l( x |  q )}{ \sqrt{ (q;q)_l}} , \hspace{20mm} \psi_0(x|q)  = \sqrt{(q;q)_\infty\over 2 \pi }  |(   e^{2i \theta};q)_\infty|
\end{equation}

\bibliographystyle{JHEP}
\bibliography{bibl}
   
  \end{document}